\documentclass[twocolumn,letter,traditabstract]{aa}
\usepackage{natbib}
\newcommand{\bibfont}{\tiny}
\usepackage{graphicx}
\usepackage[T1]{fontenc}
\usepackage{amsmath}
\usepackage{amssymb}
\usepackage{amsmath}
\usepackage{graphicx}
\usepackage{amsmath}
\usepackage{amssymb}
\usepackage{amsfonts}
\usepackage{graphicx,epsfig,graphics}
\usepackage{aas_macros}
\usepackage{graphics}
\usepackage{color}
\usepackage{epstopdf}

\def\be{\begin{eqnarray}}
\def\ee{\end{eqnarray}}
\graphicspath{{}}

\title{WMAP 9-year CMB estimation using sparsity}
\author{ \hspace{0.25in} J. Bobin\thanks{jerome.bobin@cea.fr} \and  F. Sureau \and P. Paykari \and A. Rassat \and S. Basak \and J. -L. Starck}
\institute{Laboratoire AIM, UMR CEA-CNRS-Paris 7, Irfu, SAp/SEDI, Service d'Astrophysique, CEA Saclay, F-91191 GIF-SUR-YVETTE CEDEX, France.}

\begin{document}
 
\abstract{Recovering the Cosmic Microwave Background (CMB) from WMAP data requires galactic foreground emissions to be accurately separated out. 
%Best
Most component separation techniques rely on second order statistics such as Internal Linear Combination (ILC) techniques. In this paper, we present a new 
WMAP 9-year CMB map, with 15 arcmin resolution, which is reconstructed using a recently introduced sparse component separation technique, coined Local Generalized Morphological Component Analysis (LGMCA). 
LGMCA emphasizes on the sparsity of the components to be retrieved in the wavelet domain.  
We show that although derived from a radically different separation criterion ({\it i.e.} sparsity), the LGMCA-WMAP 9 map and its power spectrum are fully consistent with their more recent estimates from WMAP 9.}

% The proposed approach yields a CMB map at 15 arcmin resolution and a power spectrum that 
%  was compared to the official WMAP 9 products~: i) the CMB temperature power spectrum is compatible with the WMAP 9 best fit to within $2\,\sigma$ error bars, ii) the higher-order statistics in the wavelet domain do not show any statistically significant evidence for non-Gaussianities in both maps at 1 degree, iii) up to the noise level, both the LGMCA and the WMAP 9 maps do not show any foreground contamination and they look similar outside the galactic center at the $1$ degree resolution. Although derived from a radically different separation criterion ({\it i.e.} sparsity), it is a remarkable fact that a CMB temperature map and its power spectrum are fully consistent with their more recent estimates from WMAP 9.}

%%%%%%%%%%%%%%%%%%%%%%%%%%%%%%%%%%%%%%%%%%%%%%%%%%%%%%%%%%%%%%%%%%
\date{Received -; accepted -}
\maketitle

%%%%%%%%%%%%%%%%%%%%%%%%%%%%%%%%%%%%%%%%%%%%%%%%%%%%%%%%%%%%%%%%%%
\section{Introduction}
The Cosmic Microwave Background (CMB) is a snapshot of the state of the Universe at the time of recombination. It provides information about the primordial Universe and its evolution to the current state. Our current understanding of our Universe is heavily based on measurements of the CMB radiation. The statistical properties of CMB fluctuations depend on the primordial perturbations from which they arose, as well as on the subsequent evolution of the Universe as a whole.  For cosmological models in which initial perturbations are of a Gaussian nature, the information carried by CMB anisotropies can completely be characterized by their angular power spectrum which depends on the cosmological parameters. This makes the precise measurement of the CMB power spectrum a gold mine for understanding and describing the Universe throughout its history.

% In the last two decades, there has been tremendous improvements both in terms of sensitivity and the angular resolution of the instruments. The ground-based, balloon-borne and satellite instruments such as WMAP \citep{2003ApJS..148....1B}, have measured the CMB anisotropies at various angular scales and at various wavelengths. The CMB angular power spectrum measured by these experiments are in remarkable agreement, and strongly support the so-called $\Lambda$-CDM (cold dark matter) model \citep{2011ApJS..192...17B,2011ApJS..192...18K}. Among these data sets, WMAP $9$-year data \citep{WMAP9_1} is the best available full sky observation, in the $23-94$ GHz frequency range.  This mission has already allowed a determination of several central cosmological parameters with great accuracy. The next full sky mission is Planck \citep{planck}, which is expected to improve the constraints on the cosmological parameters. 

In the estimation of the CMB map, the astrophysical foreground emissions from our galaxy and the extragalactic sources have to be removed. In addition, the instrumental noise hinders the estimation of the CMB map. In the low frequency regime (below ~100GHz, i.e. for WMAP channels) the strongest contamination comes from the galactic synchrotron and free-free emission, with the highest contribution at large angular scales. At higher frequencies, dust emissions dominate whereas the synchrotron and free-free emissions are low. The spinning dust is an extra emission which spatially correlates with dust and dominates at low frequencies.

Since second-order statistics provide sufficient statistics for a gaussian CMB field, most of the component separation techniques, such as the Internal Linear Combination (ILC), are built upon them to recover the CMB map from the observed sky maps. However, these techniques are not optimal for non-stationary and non-Gaussian components such as the foregrounds (or even non-stationary noise). 
On the contrary sparsity-based source separation techniques that focus on the higher-order statistics of the components have proven to be highly efficient \citep{bobin07a,2012arXiv1206.1773B}.

%Since second-order statistics provide sufficient statistics for CMB fields, most of the component separation techniques, such as Internal Linear Combination (ILC), in cosmology, relies on second order statistics to recover CMB map from the observed sky maps. However, these techniques are not optimal for non-stationary and non-Gaussian components such as foregrounds. This is where the techniques based on sparsity come to have a great impact to solve these problems. They are powerful tools because they are not only sensitive to second order statistics, but also to the higher-order statistics of the components.

%If only a few coefficients of a fixed dictionary of wavelets are enough to completely represent a given component, that particular component is said to be sparse in that dictionary.

% [JLS modif] 
%We will describe this method in detail and will apply the LGMCA method to the WMAP 9-year data.\\
%LGMCA has been evaluated on simulated Planck data in \citep{2012arXiv1206.1773B}. Its application to WMAP data is validated in \ref{sec:appendix} where LGMCA and ILC are compared on simulated WMAP 9-year data. These simulations show that LGMCA estimates a cleaner CMB map with less remaining foregrounds. However, foreground residual remain at a level which is either much lower than CMB at large scales or noise at intermediate and small scales.
%J'AI TRANSFORME LA SECTION* WMAP9 en SECTION RESULTS 
In this paper, we present a new WMAP 9-year CMB estimation based on this sparsity concept, and we compare the results to the official WMAP products.
Next section \ref{sec:compsep} briefly describes the LGMCA method.  We then describe in section \ref{sec:mapest} the processing of WMAP data and the derived LGMCA products 
are displayed in section~\ref{sec:results}. Results on WMAP simulations are also presented in Appendix~\ref{sec:appendix}.

%%%%%%%%%%%%%%%%%%%%%%%%%%%%%%%%%%%%%%%%%%%%%%%%%%%%%%%%%%%%%%%%%%
\section{\label{sec:compsep} Component Separation for CMB maps}

Exploiting the fact that foreground components are sparse in the wavelet domain (i.e. a few wavelet coefficients are enough to represent most of the energy of the component), 
LGMCA \citep{2012arXiv1206.1773B} estimates both the components of interest and the mixing matrix by maximizing the {{ level sparsity of each component; it seeks the sparsest sources possible in a wavelet basis.}} The assumption is that the observed sky is a linear combination of all the components, each resulting from a completely different physical process, and the instrumental noise.  The separation principle in this method relies on the different spatial morphologies or structures of the various foregrounds, which translate into different sparsity patterns when transformed to a fixed wavelet dictionary. A linear combination of these components decreases the level of sparsity. Therefore, reconstructing each component from the observed map, by maximizing its sparsity in wavelet space, is an efficient strategy to distinguish between physically different sources. As mentioned in Appendix~\ref{sec:appendixGMCA},  channels resolution variation and spatial variations of component emissions 
such as dust are taken into account by estimating a mixing matrix per wavelet scale and per area. Full details are given in \citet{2012arXiv1206.1773B}

In Appendix~\ref{sec:appendix}, LGMCA is evaluated on simulated WMAP data and it is found that:
 \begin{itemize}
\item The recovered power spectrum from the LGMCA map at $15$ arcminutes is within the $2\,\sigma$ error bars from the input CMB power spectrum.
\item Propagated noise in the map is the main residual contamination for LGMCA ; for low multipoles, the contaminants are much lower than cosmic variance.
\item Compared to a pixel-based localized ILC computed at 1 degree, both noise and foregrounds residuals are lower using LGMCA.
\item No significant non-gaussianities at various scales and positions are detected in LGMCA maps, either at $1$ degree or $15$ arcminutes. Compared to the pixel-based localized ILC, no significant difference is observed for LGMCA at 1 degree compared to the errors expected.
\end{itemize}
Altogether, these results incentivize applying LGMCA to  WMAP9 data.

% There also exists other ILC-based techniques, such as Needlet-ILC, which has been applied to WMAP 7-year data \citep{2012MNRAS.419.1163B}. The Needlet-ILC estimates a minimum-variance weight for each %ILC parameters 
% frequency channel following a local and multi-scale procedure. As shown in \citep{2012arXiv1206.1773B}, sparsity in conjunction with the refined local estimation of the mixing matrices, allows for an improved estimation of the CMB map.\\
%More specifically, the WMAP data are contaminated by thermal dust emissions especially in the $W$ band at $94$ GHz. However, the WMAP data alone do not offer enough information for a clean separation of this component. Ancillary data was therefore added to the component separation process to fully take benefit of the WMAP data. In the following, the LGMCA will also be performed on the Finkbeiner dust map and the H$\alpha$ map (as a tracer of the free-free emission) in addition to the WMAP data. 

%%%%%%%%%%%%%%%%%%%%%%%%%%%%%%%%%%%%%%%%%%%%%%%%%%%%%%%%%%%%%%%%%%
\section{Map and Power Spectrum Estimation}
\label{sec:mapest}

\begin{figure}[htb]
\centering{
\includegraphics[scale=0.12]{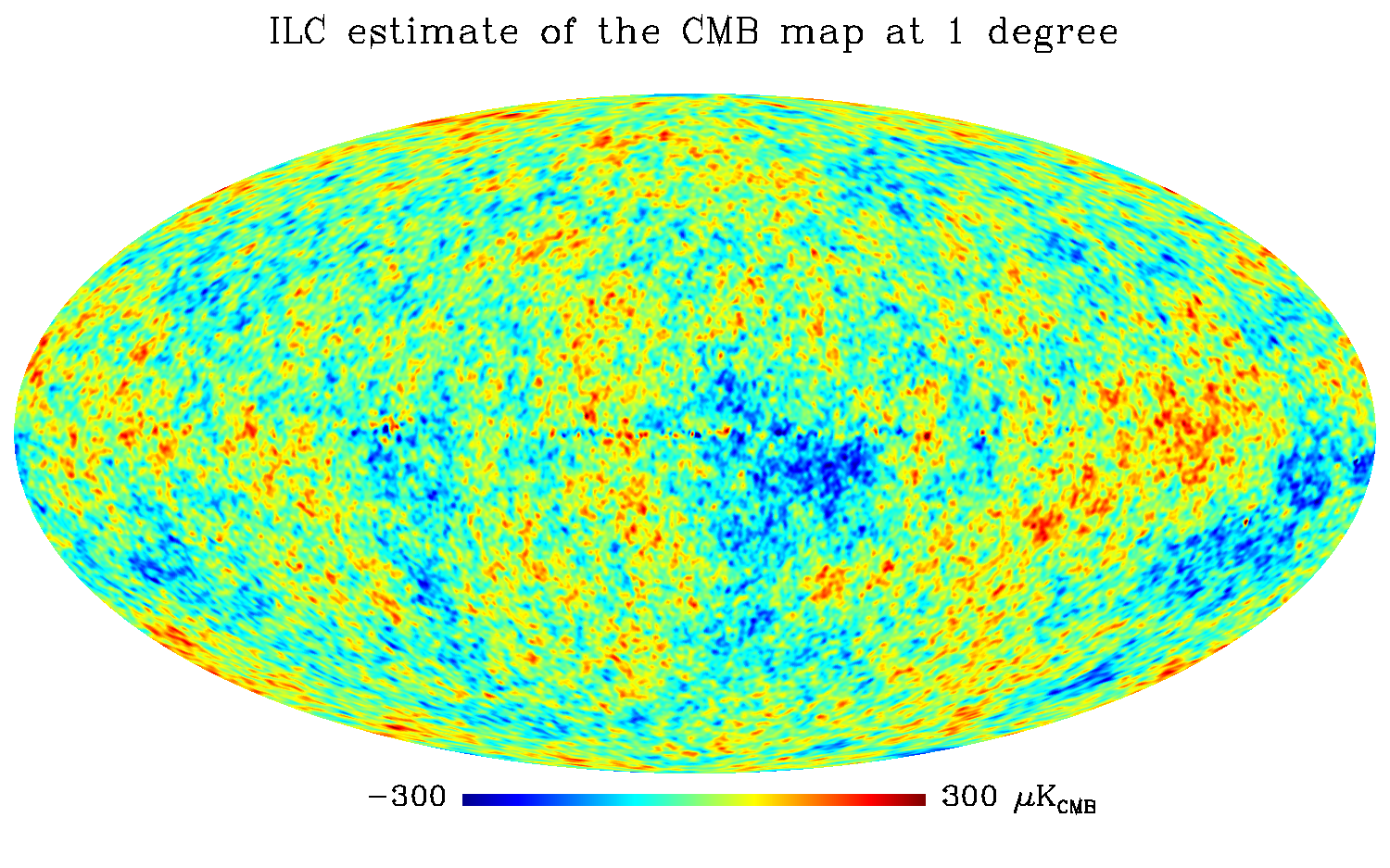}
\caption{ILC  official WMAP 9 years map ($1$ degree resolution). Units in $\mu K$.}
\label{fig:wmap9}
}
\end{figure}

The WMAP satellite has observed the sky in five frequency bands denoted K, Ka, Q, V and W centered on $23$, $33$, $41$, $61$ and $94$GHz respectively. The released data includes sky maps obtained with ten differencing assemblies, for nine individual years; per year, there is one map for the K band and one for the Ka band, two for the Q band, two for the V band and four for the W band. These maps are sampled using the HEALPix pixelization scheme at a resolution corresponding to {\it nside} of $1024$.  
% These maps contain a mixture of astrophysical foreground emissions as well as the instrumental noise, superimposed on CMB, which degrade the attainable accuracy of cosmological information. The Galactic foregrounds are particularly significant on large angular scales. Hence, removal of foreground emissions is the primary step towards accurate estimation of angular power spectrum of CMB.  
At first, we average all the differencing-assembly maps obtained for the same frequency band, which yields five band-averaged maps. However, these maps are not offset-corrected. In order to determine the offset value for a particular frequency band, we use the standard resolution of the $9$-year band-average maps of WMAP as reference maps. Offset values are obtained by determining the mean of the difference between the band-averaged map being considered, and the reference map. % We then implement LGMCA on these offset-corrected maps. 

At the WMAP frequencies, the major sources of contamination in the maps are the synchrotron, free-free, spinning dust and thermal dust. 
To model the foreground contamination, we use two foreground templates in our analysis: dust at $100$ microns, as obtained by \citep{1998ApJ...500..525S} and the composite all-sky H-alpha map of \citep{2003ApJS..146..407F}. Among these templates, the thermal dust template is the most important one, as it helps in removing the dust emission on small scales, which is otherwise significant in the W channel. As the spinning dust is spatially correlated with thermal dust, the thermal dust template also helps in reducing the spinning dust residuals. It is usually standard to use the $408$ MHz synchrotron map of \citep{1981A&A...100..209H} as a template for the synchrotron emission. However, this map has quite a low resolution of about $1$ degree. Furthermore, adding this template to LGMCA did not show any improvement for the component separation.

{\em \bf LGMCA map:} following \citep{2012arXiv1206.1773B}, the LGMCA is applied to the five offset-corrected WMAP maps and the two templates used as extra observations. 
Details on LGMCA parameters are given in Appendix~\ref{sec:appendixWMAPGMCA}. %  Figure~\ref{fig:real_maps} shows the estimated LGMCA CMB at  $15$ arcmin resolution  and $1$ degree. 

{\em \bf LGMCA power spectrum:} using the mixing matrices  previously estimated from all data, we can estimate a CMB map from each of the nine individual year data set. 
From these nine maps, we have derived all $36$ possible cross-spectra using the { high resolution} temperature analysis mask kq85 with $(f_{sky}=0.75) $\footnote{{\it http://lambda.gsfc.nasa.gov/product/map/dr5/m\_products.cfm}} provided by WMAP collaboration.  The final CMB power spectrum is then obtained by averaging them and by using a MASTER mask deconvolution  \citep{2002ApJ...567....2H}.
In contrast to the CMB and foreground signals, noise is uncorrelated between different years of data and  it is therefore considerably reduced in the averaged cross  spectrum.

\begin{figure*}[htb]
\centering{
\includegraphics[scale=0.12]{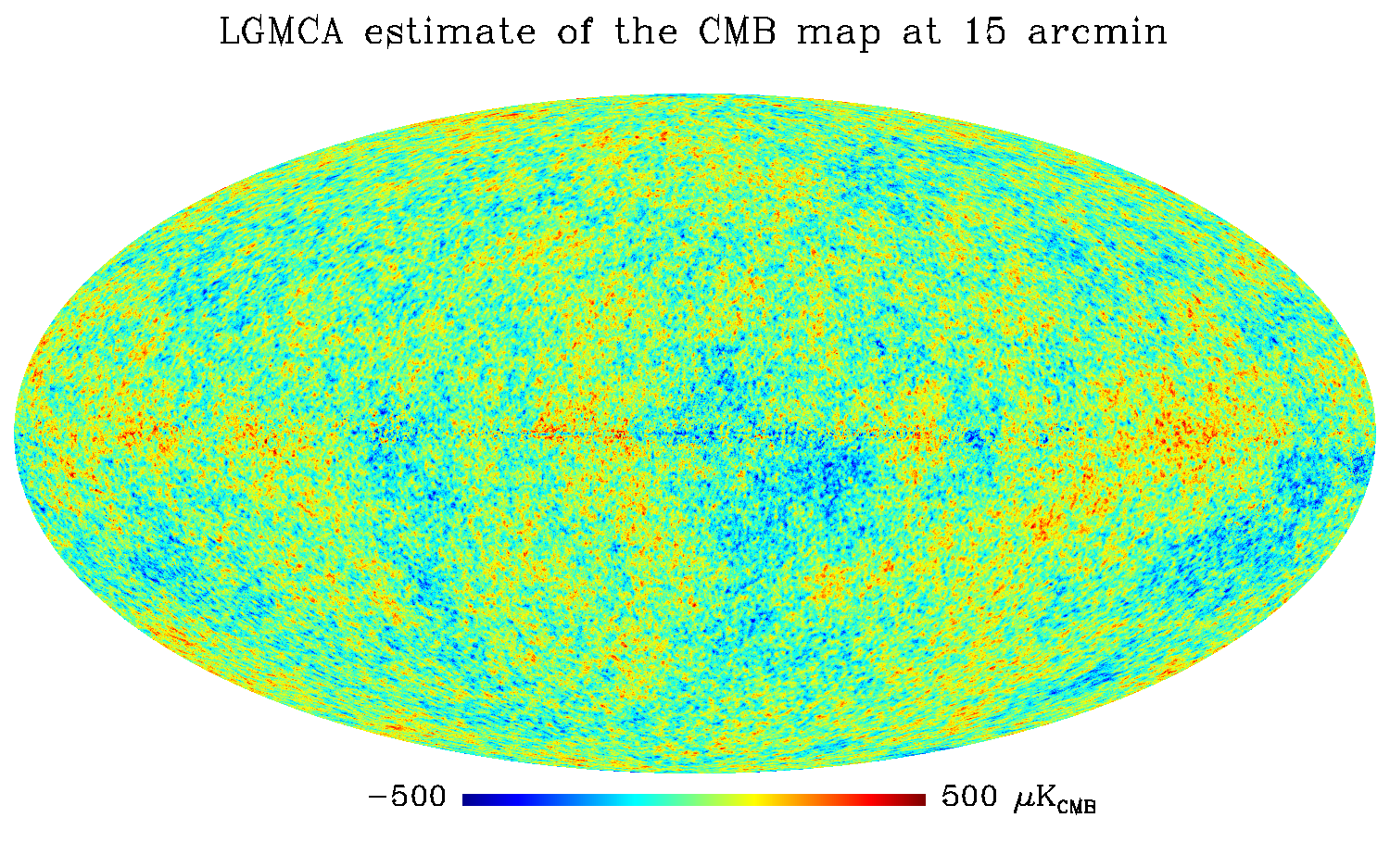}
\includegraphics[scale=0.12]{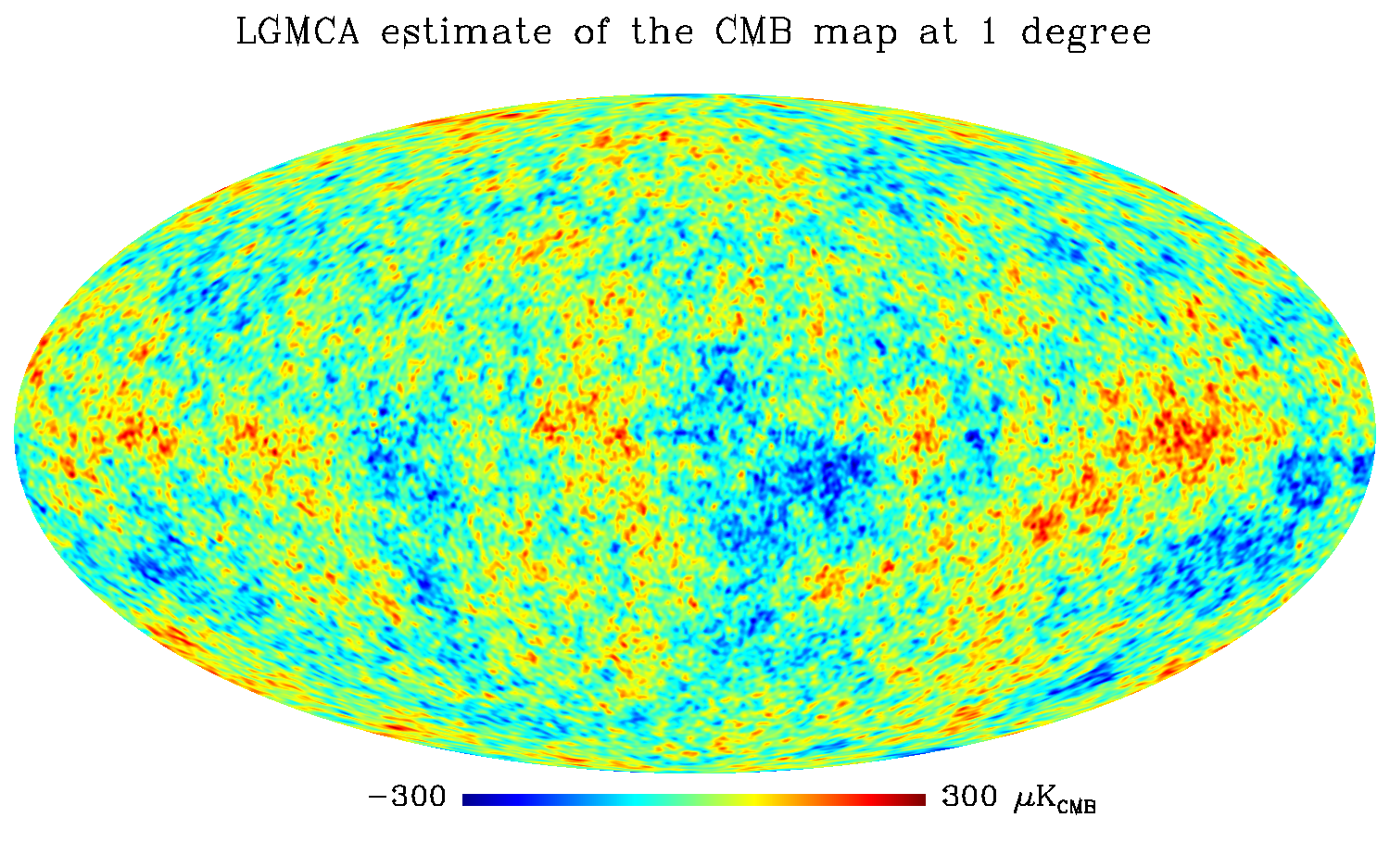}
\caption{Estimated LGMCA CMB map from WMAP ($9$ years)  at $15$ arcmin resolution  and $1$ degree. Units in $\mu K$.}
\label{fig:real_maps}
}
\end{figure*}

\begin{figure*}[htb]
\centering{
\includegraphics[scale=0.12]{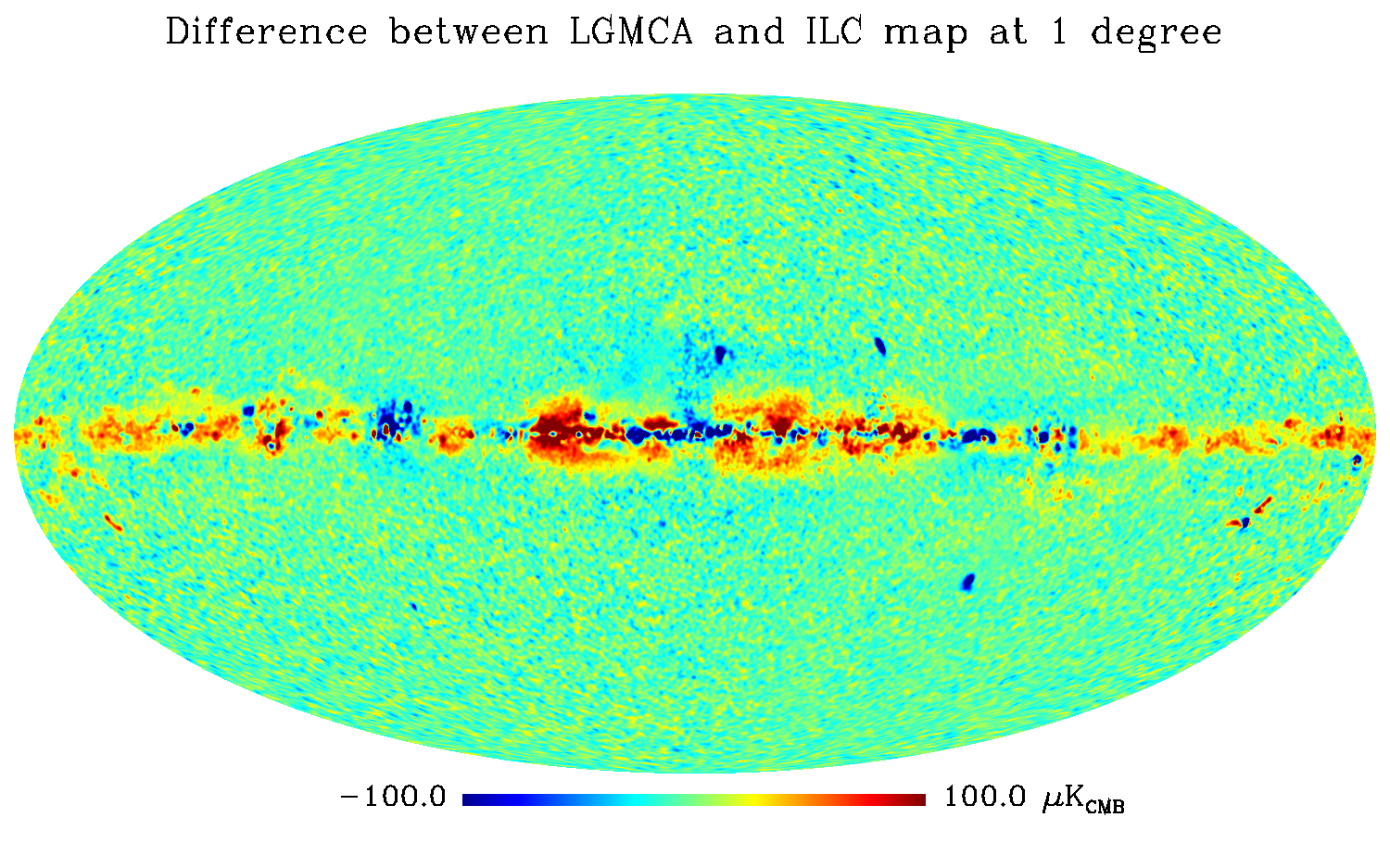}
\includegraphics[scale=0.12]{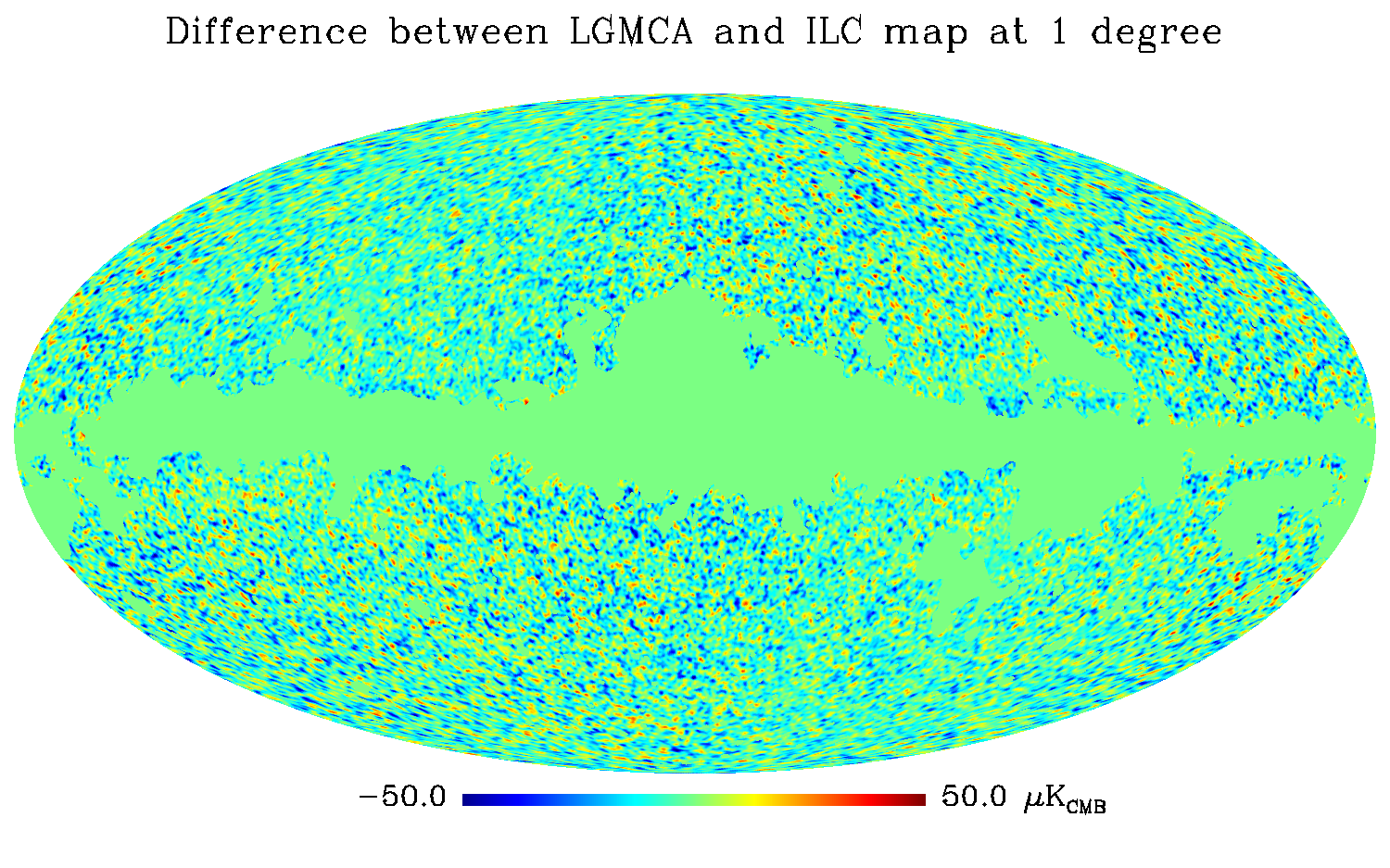}
}
\caption{Difference between the estimated CMB map with LGMCA and the official ILC map. The bottom panel shows the same difference map masked with the kq85 mask.}
\label{diffs_wmap}
\end{figure*}

%%%%%%%%%%%%%%%%%%%%%%%%%%%%%%%%%%%%%%%%%%%%%%%%%%%%%%%%%%%%%%%%%%
% [JLS modif] - Actually he did not ask, but I did for the description at the end of the intro that he asked] 
\section{\label{sec:results} Results}

\subsection*{LGMCA Map and Power Spectrum}

% Figure~\ref{fig:real_maps} shows from top to bottom the official ILC  WMAP 9 years map which has a $1$ degree resolution, the GMCA map at the same resolution and 
% the GMCA at full resolution ($15$ armin).   
Figure~\ref{fig:wmap9} shows the official ILC WMAP 9 years CMB map, which has a $1$ degree resolution, and
Figure~\ref{fig:real_maps} shows the LGMCA WMAP CMB map at $15$ arcmin resolution  and at $1$ degree.
The $15$ arcmin LGMCA map exhibits slight high-frequency structures which are likely related to point sources emission on the galactic center. 
The two  $1$ degree maps look very clean, even on the galactic center.
Figure~\ref{diffs_wmap} features the difference at resolution $1$ degree between the CMB map estimated by LGMCA and the official ILC-based CMB map provided by the WMAP consortium. 
The difference map shows significant foreground residuals on the galactic center, that are however roughly three times lower than the CMB level. 
% It is therefore very difficult to identify which map is more contaminated on the galactic center. 
When masking the galactic center with the  kq85 WMAP mask (Fsky=$75 \% $), no significant feature can be seen anymore.
 
\begin{figure*}[htb]
\centering{
\includegraphics[scale=0.2]{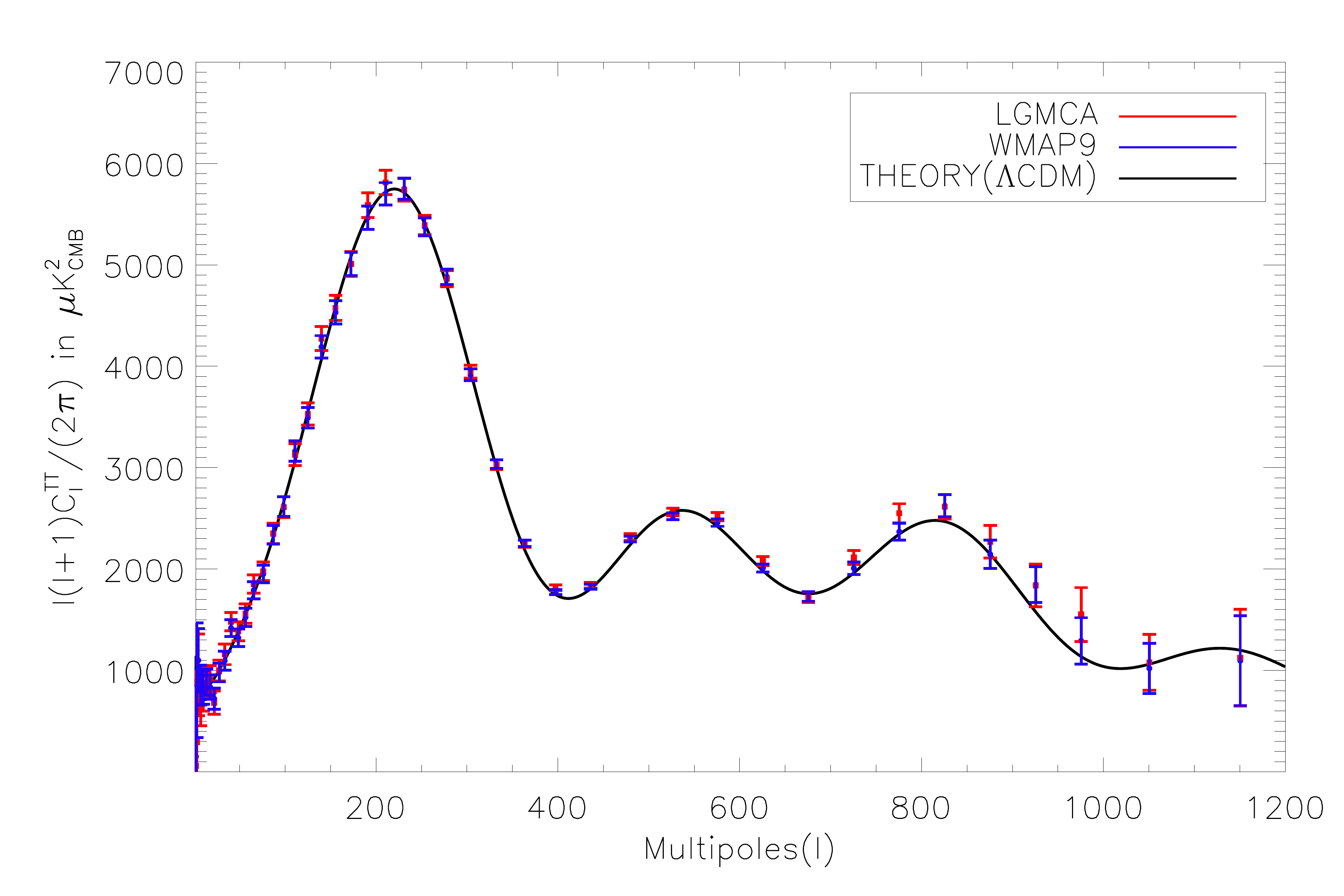}
\includegraphics[scale=0.2]{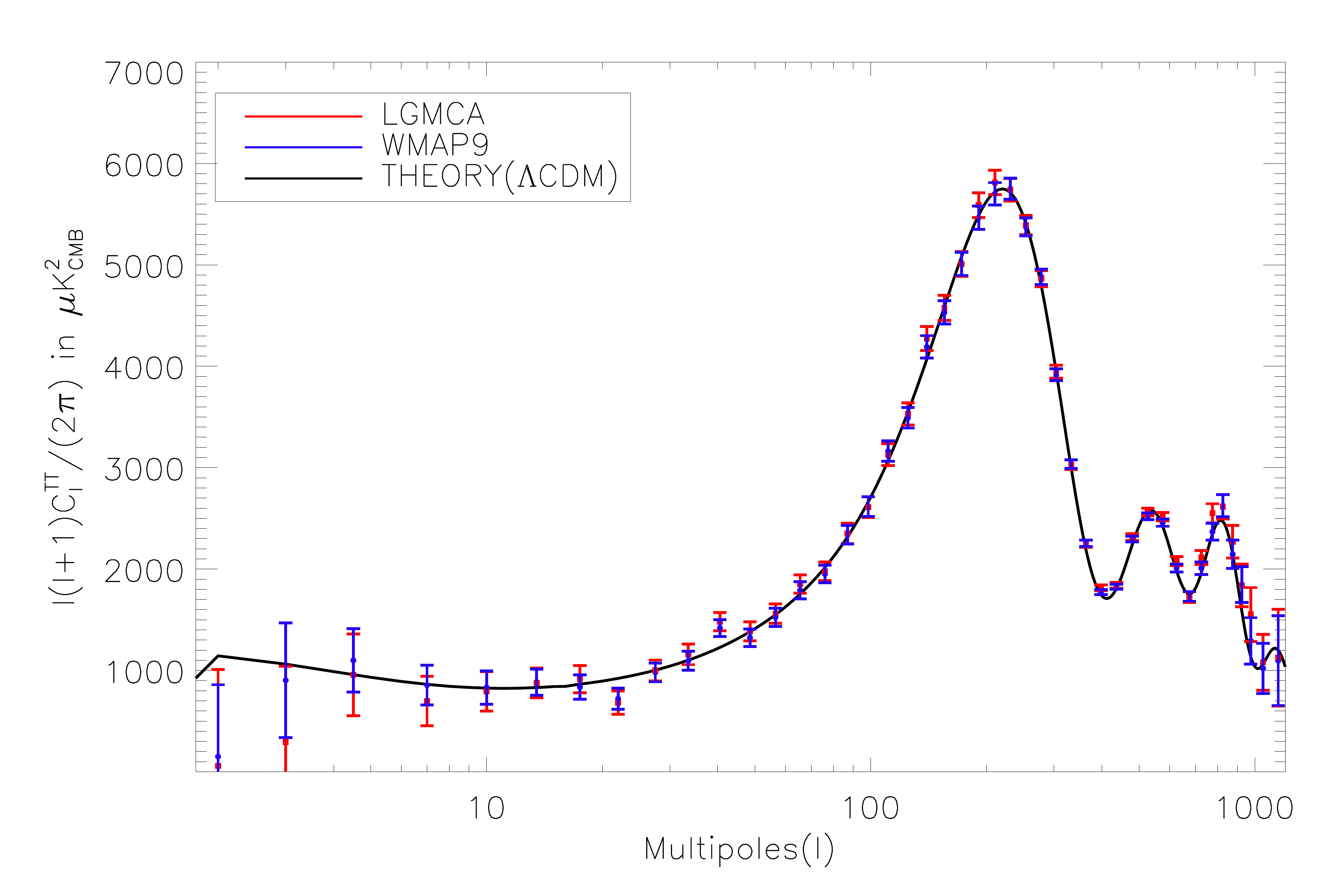}
\caption{Estimated CMB map power spectrum from WMAP ($9$ years) in linear scale (top panel) and logarithmic scale (bottom panel). Units in $\mu K^2$.}
\label{fig:real_ps_1}
}
\end{figure*}

%  The power spectrum is estimated from the $15$ arcmin LGMCA map using the same kp85 mask and the MASTER method \citep{2002ApJ...567....2H}. 
The estimated power spectrum is displayed in Figure~\ref{fig:real_ps_1}. 
The errors bars, essentially the cosmic variance and the noise-related variance, are derived from classical power spectrum variance estimators as described in \citep{SupplWMAP9}. 
Note that except for $\ell < 32$, the official WMAP 9 years power spectrum (in blue in Fig.~\ref{fig:real_ps_1}) is computed from the V and W bands only. 
In opposite, the spectrum in red in Fig.~\ref{fig:real_ps_1} is derived from the full dataset. Even if the estimation procedure differs slightly, %  from the one the WMAP consortium adopted, 
it is remarkable that the two power spectra look very similar. They also tend to depart from each other with a slightly lower third multipole. As well the measurement about the third peak seem to be higher to some extent. However both spectra are compatible at all scales within $2\,\sigma$ error bars. 

% \subsection*{Sanity check}
\begin{figure}[htb]
\includegraphics[scale=0.2]{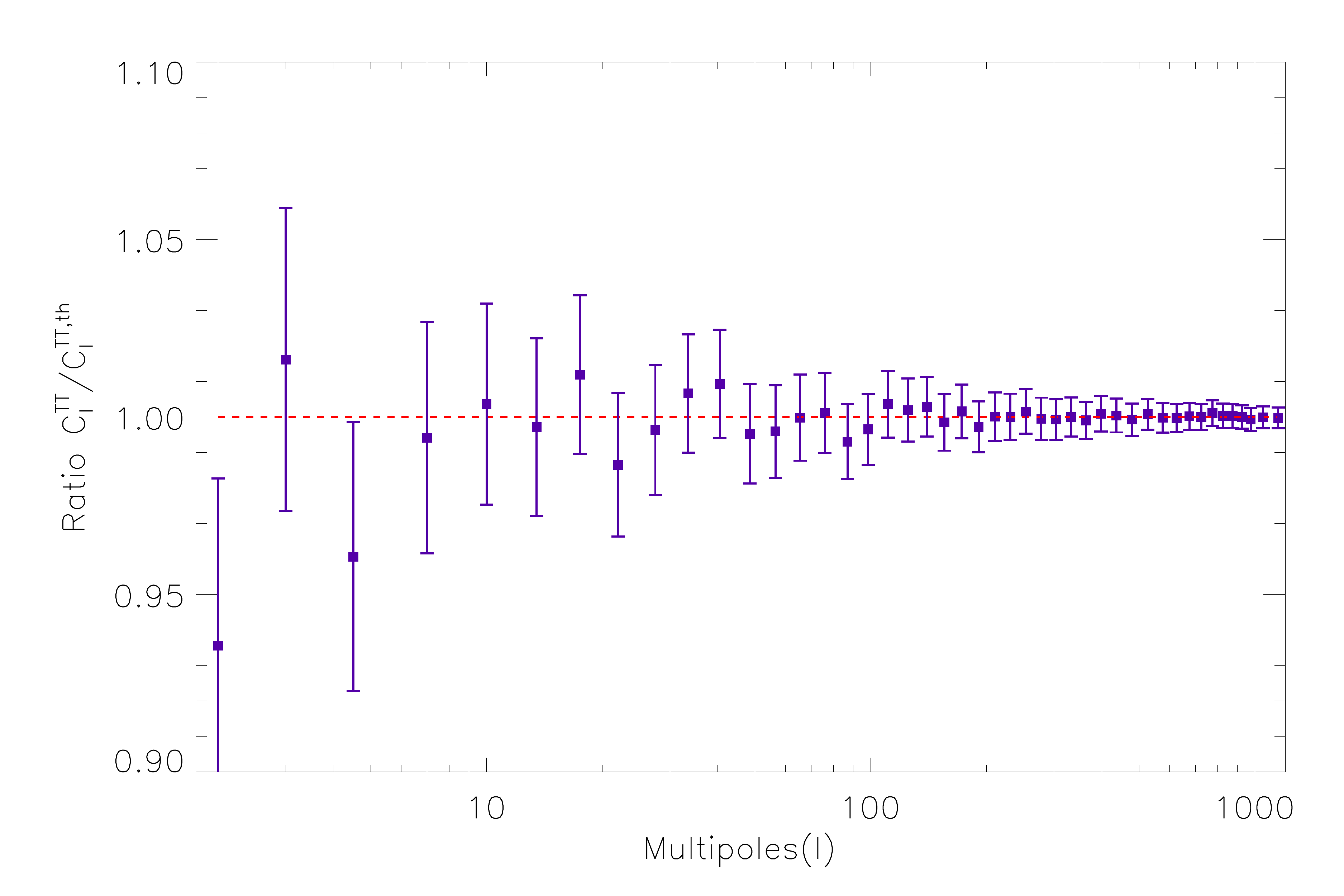}
\caption{Mean ratio of $C_\ell^{TT}/C_\ell^{TT,th}$ over $100$ random CMB simulations. The error bars are related to the cosmic variance over the $100$ simulations.}
\label{fig:check_ps_1}
\end{figure}

\begin{figure*}[htb]
\begin{center}
\includegraphics[scale=0.33]{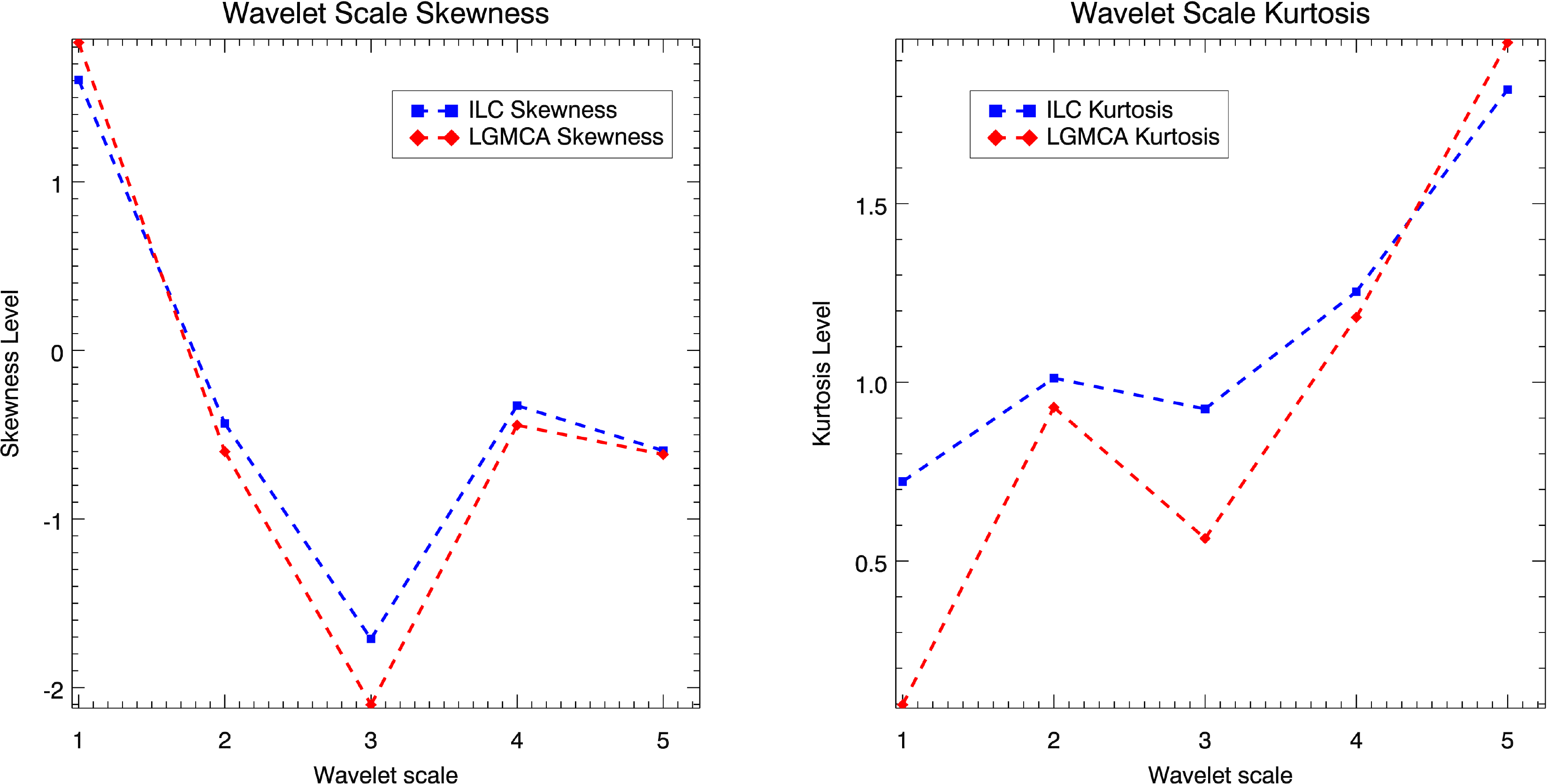}
\end{center}
\caption{Comparison of skewness and Kurtosis in ILC and LGMCA map at $1$ degree computed for the various wavelet scales described in Figure~\ref{fig:simu_hos_0} (left). The wavelet filters peaks respectively about multipole $240$, $120$, $60$, $30$ and $15$ for scales $1$ to $5$.}
% A $59 \%$ mask was used to compute these statistics (see text). }
\label{fig:data_hos_1}
\end{figure*}

{\em Sanity check:} The LGMCA algorithm computes mixture parameters ({\it i.e.} mixing matrices and their inverse) which are then applied to the data to estimate the CMB map. It is important to note that the final CMB map linearly depends on the input data. This makes it possible to check whether the inversion process may induce some bias at the level of the estimated CMB map and its power spectrum (assuming no calibration or beam errors). For that purpose, we applied exactly the same parameters we computed from the real data to $100$ random CMB realizations. These realizations were generated as Gaussian random processes with a power spectrum defined by the WMAP 9-year best-fit theoretical power spectrum. { The point spread function of these simulations are chosen as the beams of the 9-year band-average maps of WMAP thus providing pure CMB simulations that mimic the CMB part of the WMAP 9 years data.} Figure~\ref{fig:check_ps_1} shows in blue the ratio between the estimated and theoretical power spectrum $C_\ell^{TT}/C_\ell^{TT,th}$ computed from an average of $100$ random CMB simulations. { The theoretical power spectrum $C_\ell^{TT,th}$ of the simulated CMB maps is exactly defined as the WMAP 9-year best-fit theoretical power spectrum.} The error bars are defined as the cosmic variance from $100$ simulations. If the estimation by LGMCA of the CMB map and more specifically its power spectrum were biased, the $C_\ell^{TT}/C_\ell^{TT,th}$ should depart from $1$. Figure~\ref{fig:check_ps_1} shows that there is no statistical evidence of discrepancy from $1$. We therefore conclude that the LGMCA does not introduce any bias at the level of the CMB power spectrum. { Any bias of the estimated CMB map will come from remaining noise and foreground contamination}. \\
%This does not mean that the estimated is unbiased as bias can originate from the calibration errors or beam errors which would impact any estimation process anyway.
% \subsection*{Higher Order Statistics --- Non-Gaussianities}
{\em Higher Order Statistics --- Non-Gaussianities:} 
Higher order statistics were also computed for the full $9$ year CMB maps recovered at $1$ degree by ILC and LGMCA to assess potential differences in their distribution. 
% Note that the official ILC map includes a post-processing step subtracting the "ILC-bias" arising from remaining foreground contamination by estimating foreground templates from the data, resulting in a non-linear process, which would lead to a much more complex estimation of error bars on the statistics. However, the difference between the two maps is expected to be very low in the region where these statistics where computed (see below).
% As performed for power spectra estimation,  
The $75 \%$ mask was employed on both maps  to avoid computing the higher order statistics in regions contaminated by foreground residuals. 
Sparse inpainting was then performed to interpolate the signal inside the mask \citep{Starck2013} in order to avoid artifacts on the wavelet coefficients, and the 
skewness and kurtosis were calculated for each wavelet scale considering only wavelet coefficients within the mask.
% To take into account this step after smoothing to a $1$ degree map, an analysis mask was derived by convolving the mask with the appropriate gaussian kernel (from $15$ arcminutes to $1$ degree) and setting to zero any pixel below $0.99$. This results in a mask covering about $59 \%$ of the sky.
% Higher order statistics were computed in this mask for different wavelet scales and latitude bands (using the filters displayed in Figure \ref{fig:simu_hos_0}). 
These statistics were centered and normalized by processing similarly a set of $100$ realizations of noise and CMB according to the $\Lambda$-CDM fit provided by the WMAP consortium.
 Figures \ref{fig:data_hos_1} shows the skewness and kurtosis versus the wavelet scale, and illustrates that not only both ILC and LGMCA maps are compatible with no non-gaussianities,
 % (except for the cold spot displaying a high kurtosis in both maps at the last wavelet scale), 
 but also that no significant difference between ILC and LGMCA can be found with these statistics at that resolution.

%\begin{figure*}[!h]
%\begin{center}
%\includegraphics[scale=0.4]{WMAP9_comparison_ILC_LGMCA_data_kp85_mask512_1deg_kurtskew.pdf}
%\end{center}
%\caption{Comparison of skewness and kurtosis in ILC and LGMCA map at 1 degree computed for various wavelet scale. A 59\% mask was used to compute these statistics (see text). Error bars are plotted as $\pm 1 \sigma$ around the computed value, estimated from a set of 100 simulations of CMB and Noise.}
%\label{fig:data_hos_1}
%\end{figure*}

%%%%%%%%%%%%%%%%%%%%%%%%%%%%%%%%%%%%%%%%%%%%%%%%%%%%%%%%%%%%%%%%%%
\section{Conclusion}
We have investigated how sparsity could be used for WMAP CMB map reconstruction.
Based on WMAP simulations, we have shown that LGMCA provides a low foreground map, and that noise remains the major source of contamination.
Then a high resolution ($15$ arcmin) clean CMB map has been computed from the full WMAP 9 years dataset, and its power spectrum was estimated. 
% This component separation method radically departs from standard methods in astrophysics such as ILC-based approaches as it emphasizes the sparsity of the components to be retrieved rather than focusing on second-order statistics as separation criterion. 
Remarkably, though LGMCA-based and official WMAP 9 years power spectrum were derived from completely different estimation procedures, they are in very good agreement and compatible within $2 \, \sigma$ error bars. 
Lastly, non-gaussianity tests based on higher order statistics were carried out, and do not show statistically significant departure from gaussianity at resolution $1$ degree.
% degree except the cold spot. 
% [JLS modif] Do not think we cannot speak about the differences in the galactic center. 
The LGMCA and the official WMAP 9 maps essentially differ close to the galactic center where it remains extremely difficult to assess which map is less contaminated by foreground residuals or biases due to chance correlations in between CMB and foregrounds.

The LGMCA code is available at {\it http://www.cosmostat.org/lgmca} and the LGMCA CMB map as well as the estimated power spectrum are available at {\it http://www.cosmostat.org/product}.

\section*{Acknowledgement}
This work was supported by the European Research Council grant SparseAstro (ERC-228261).

 \bibliographystyle{aa} 
\bibliography{lgmca_wmap9}

\clearpage
\newpage

%%%%%%%%%%%%%%%%%%%%%%%%%%%%%%%%%%%%%%%%%%%%%%%%%%%%%%%%%%%%%%%%%%
\appendix

\section{\label{sec:appendixGMCA} The LGMCA Method}

\subsection*{The GMCA Framework}
The GMCA (Generalized Morphological Component Analysis) method is based on 
blind source separation (BSS) \citep{2012arXiv1206.1773B}. In the framework of BSS, each of the five WMAP frequency channels are modeled as a linear combination of $n$ components:
\begin{equation}
\forall i=1,\cdots,5; \, x_i = \sum_{j=1}^n a_{ij} s_j + n_i
\end{equation}
where $s_j$ stands for the $j$-th component, $a_{ij}$ is a scalar that models for the contribution of the $j$-th component to channel $i$ and $n_i$ models the instrumental noise. This problem is more conveniently recast into the matrix formulation~:
\begin{equation}
{\bf X} = {\bf A S} + {\bf N}
\end{equation}
In practice, the number of components is set to $n=5$ which allows for more degrees of freedom to get a clean CMB map while keeping $\bf A$ invertible. Contrary to standard approaches in astrophysics (see \citep{2012arXiv1206.1773B} and references therein), the GMCA relies on the sparsity of the components $\bf S$ in the wavelet domain. Taking the data to the wavelet representation only alters the statistical distribution of the 
data coefficients without affecting its information content. A wavelet transform tends to grab the informative coherence between pixels while averaging the noise contributions, thus enhancing the structure in the data. This allows to better distinguish components that do not share the same sparse distribution in the wavelet domain. In addition, sparsity has the ability to be more sensitive to non-Gaussian processes, which has been shown to improve the foreground separation method.

Having $\boldsymbol{A}$ as the mixing matrix and $\boldsymbol{\Phi}$
as a wavelet transform, we assume that each source $s_{j}$ can be %represented as a 
sparsely represented in ${\bf \Phi}$; $s_{j}=\alpha_{j}\boldsymbol{\Phi}$. The multichannel noiseless data $\boldsymbol{Y}$ can be written as
\begin{equation}
\boldsymbol{Y}=\boldsymbol{A}\mathbf{\alpha}\boldsymbol{\Phi}\:.\label{eq:tensor1-1}\end{equation}
where $\mathbf{\alpha}$ is a $N_{s}\times T$ matrix whose rows are $\alpha_{j}$.

This means the sparsity of the sources in $\mathbf{\Phi}$ translates
into sparsity of the multichannel data $\boldsymbol{Y}$. The GMCA algorithm seeks an unmixing scheme through the estimation of $\boldsymbol{A}$, which leads to the sparsest sources $\boldsymbol{S}$. This is expressed by the following optimization problem (written in the augmented Lagrangian
form)
\begin{equation}
\min\frac{1}{2}\left\Vert \boldsymbol{X}-\boldsymbol{A}\mathbf{\alpha}\boldsymbol{\Phi}\right\Vert _{F}^{2}+\lambda \left \| \mathbf{\alpha} \right \|_{p}^{p}\:,\end{equation}
where typically $p=0$ (or its relaxed convex version with $p=1$)
and ${\bf \left\Vert \boldsymbol{X}\right\Vert }_{\mathrm{F}}= \textrm{sqrt} \left(\textrm{trace}(\boldsymbol{X}^{T}\boldsymbol{X})\right)$
is the Frobenius norm.

\subsection*{Local GMCA}
The Local-GMCA (LGMCA) algorithm \citep{2012arXiv1206.1773B} has been introduced as an extension of GMCA:
\begin{itemize}
\item multi-frequency instruments generally provide observations that do not share the same resolution. For example, the WMAP frequency channels have a resolution that ranges from $13.2$ arcmin for the $W$ band to $52.8$ arcmin for the K band. The makes the linear mixture model underlying the GMCA algorithm not valid. It is customary to alleviate this issue by degrading the frequency channels down to a common resolution prior to applying any component separation technique (the official CMB map provided by the WMAP consortium has a resolution of $1$ degree). For this purpose, the data are decomposed in the wavelet domain and at each wavelet scales we only use the observations with invertible beams and then degrade the maps to a common resolution. This allows to estimate a CMB map with a resolution of $15$ arcmin.\\ 
\item most foreground emissions ({\it e.g.} thermal dust, synchrotron, free-free) have electromagnetic spectra that are not spatially constant. In the framework of GMCA, this translates into a mixing matrix $\bf A$ that also varies across pixel. Dealing with the variation across pixels of the electromagnetic spectrum of some of the components, the LGMCA estimates the mixing matrices on patches at various wavelet scales with band-dependent size.
\end{itemize}
The LGMCA algorithm has been implemented and evaluated on simulated Planck data in \citep{2012arXiv1206.1773B}. 

\section{\label{sec:appendixWMAPGMCA} LGMCA parameters for WMAP data}

\begin{figure}[htb]
\includegraphics[scale=0.2]{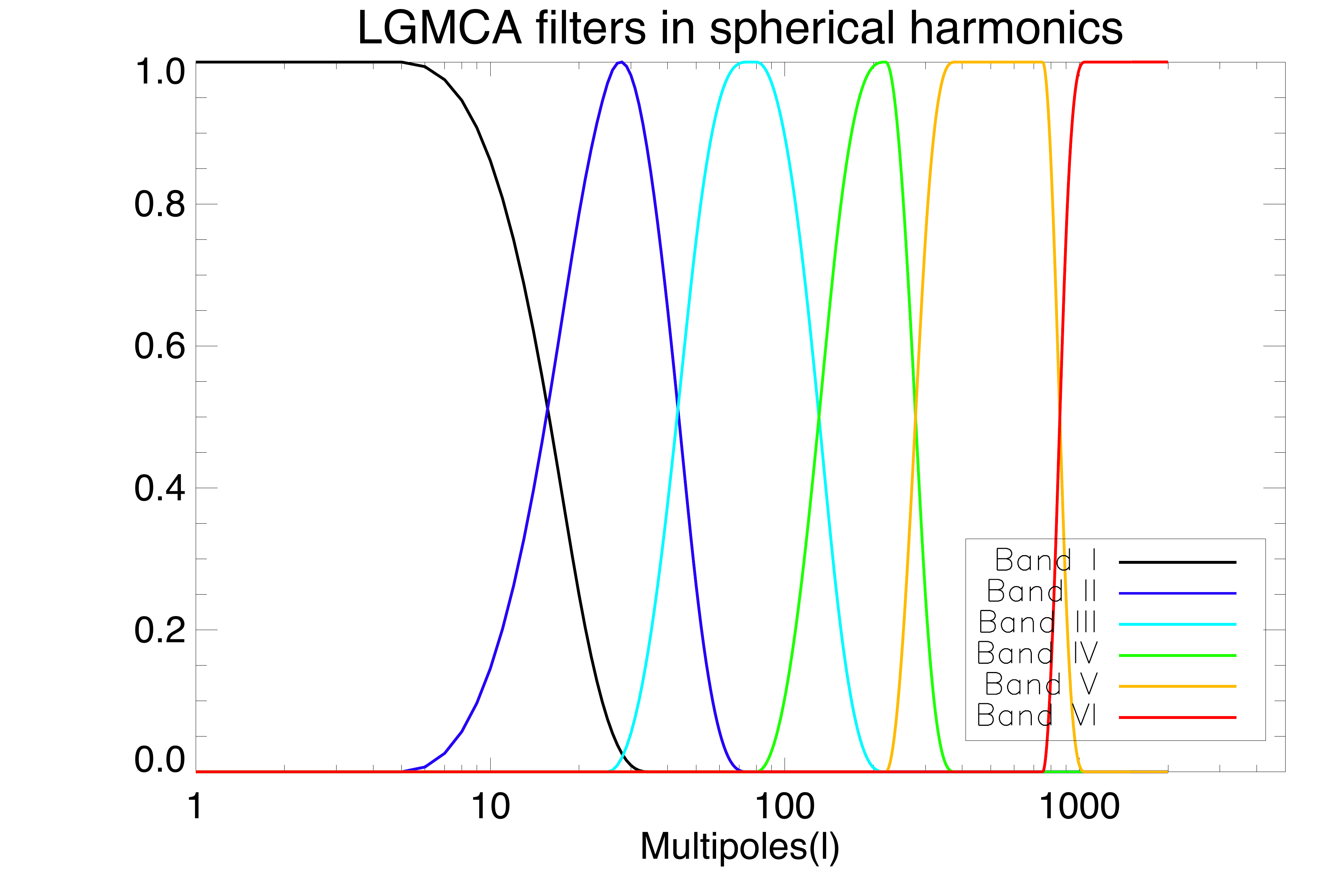}
\caption{Filters defining the wavelet bands used in LGMCA.}
\label{fig:bands}
\end{figure}

As described in Appendix~\ref{sec:appendixGMCA}, LGMCA mixing matrices are estimated from a set of input channels at a given resolution, in a patch of data at a given wavelet scale. For WMAP data, the parameters used by LGMCA to compute these matrices are described in Table ~\ref{tab_resol_planck}. Figure~\ref{fig:bands} displays the filters in spherical harmonics defining the wavelet bands where the derived weights (by inverting these mixing matrices) were applied to.

%The parameters used to process the WMAP data are described in Table ~\ref{tab_resol_planck}. LGMCA is then applied on $6$ different wavelet bands; these bands are entirely described by filters in spherical harmonics, shown in Figure~\ref{fig:bands}.
% LGMCA is applied on $6$ different wavelet bands; these bands are entirely described by filters in spherical harmonics, shown in Figure~\ref{fig:bands}.
% In each band, only a subset of the data is used and downgraded to a common resolution defined as the lowest resolution of the maps within the selected subset. % [JLS modif] Did not want to speak about the wavelet decomposition. So I modified also the following table too
%In each band, mixing matrices are estimated in the wavelet domain with a band-dependent number of scales (WT scales).
%  As described in \citep{2012arXiv1206.1773B}, the mixing matrices are estimated on square patches the size of which also depends on the band. The parameters used to process the WMAP data are described in Table ~\ref{tab_resol_planck}.

\begin{table}
\begin{center}
\vspace{0.1in}
\begin{tabular}{|c|c|c|c|}
\hline
Band & Obs. & Patch size  & Res. \\
\hline
\hline
I & WMAP 9 data  & no  & 60 \\
 & dust and H$\alpha$  &  &\\
 \hline
II & WMAP 9 data  & 256  &53 \\
 & dust  &   &\\
 \hline
III & Ka,Q,V,W bands  & 128  & 39 \\
 & dust  &   &\\
 \hline
IV & Q,V,W bands  & 64  &30 \\
 & dust  &   &\\
 \hline
V & V,W bands  & 64 &22 \\
 & dust  &   &\\
  \hline
VI & W band  & 64  & 15 \\
 & dust  &   &\\
\hline
\end{tabular}
\vspace{0.1in}
\end{center}
\caption{Parameters used in LGMCA to process the WMAP data with ancillary data. For each band, the second column gives the subset of data used to analyze the data, the third column provides the size of the square patches at the level of which the analysis is made, the fourth column gives the common resolution of the data.}
\label{tab_resol_planck}
\end{table}

\section{\label{sec:appendix} Simulations}

%Simulations provide a mean to explore the potential performances of component separation methods. In this section, the LGMCA algorithm is performed on WMAP9 simulated data. The WMAP9 data have been simulated using  the Planck Sky Model (PSM) developed by J. Delabrouille and collaborators\footnote{For more details about the PSM, we invite the reader to visit the PSM website: {\it http://www.apc.univ-paris7.fr/~delabrou/PSM/psm.html}.} in \citep{PSM12}. 
In this section, the LGMCA algorithm is applied to data simulated by the Planck Sky Model (PSM) developed by J. Delabrouille and collaborators\footnote{For more details about the PSM, we invite the reader to visit the PSM website: {\it http://www.apc.univ-paris7.fr/~delabrou/PSM/psm.html}.} in \citep{PSM12}.
The PSM models the astrophysical foregrounds in the range of frequencies probed by WMAP, the simulated instrumental noise and the beams. In detail, the simulations were obtained as follows.
\begin{itemize}
\item{\bf Frequency channels:} the simulated data are comprised of the $5$ WMAP channels at frequency $23$, $33$, $41$, $61$ and $94$ GHz. The frequency-dependent beams are perfectly isotropic PSF; their profiles have been obtained as the mean value of the beam transfer functions of at each frequency as provided by the WMAP consortium ($9$ years version).
\item{\bf Instrumental noise:} instrumental noise has been generated according to a Gaussian distribution with the covariance matrix provided by the WMAP consortium ($9$ years version).
\item{\bf Cosmic microwave background:} the CMB map is a Gaussian random realization with theoretical power spectrum defined as the WMAP (9 years) best-fit power spectrum (from the $6$ cosmological parameters model). The simulated CMB is perfectly Gaussian, and no non-Gaussianity ({\it e.g.} lensing, ISW, $\mbox{f}_{NL}$) has been added. This will allow for non Gaussianity tests under the null assumption in the sequel.
\item{\bf Dust emissions:}  the galactic dust emissions is composed of two distinct dust emissions: thermal dust and spinning dust ({\it a.k.a.} anomalous microwave emission). Thermal dust is modeled with the Finkbeiner model (\citep{Fink99}), which assumes that two hot/cold dust populations contribute to the signal in each pixel. The emission law of thermal dust varies across the sky.
\item{\bf Synchrotron emission:} The synchrotron emission, as simulated by the PSM, is an extrapolation of the Haslam $408$MHz map (\citep{Haslam:1982dk}). The emission law of the synchrotron emission is an exact power law with a spatially varying spectral index.
\item{\bf Free-free emission:} the spatial distribution %geometry 
of free-free emission is inspired by the H$\alpha$ map built from the SHASSA and WHAM surveys. The emission law is a perfect power law with a fixed spectral index.
\item{\bf Point sources:} infrared and radio sources were added based on existing catalogs at that time (including WMAP7 sources). In the following, the brightest point sources are masked prior to the evaluation results.
\end{itemize}
A simulated WMAP dataset is produced for each of the $9$ years. This allows to process the simulated data in the exact similar manner as the WMAP data are processed. 

\subsection*{Component Separation}

The same templates and parameters as listed in Table~\ref{tab_resol_planck} were used for LGMCA. We also implemented an ILC as for the WMAP9 release: first computing the weights in the same regions as for the WMAP9 release, then smoothing them to 1.5 degree and finally applying them to the data at 1 degree in the same regions as defined in the official WMAP9 product. Note that no post-processing was performed to subtract the ILC bias due to foreground propagation as in the official product. This allows to compare on the simulations the relative performance of LGMCA and a localized ILC in pixel space at a resolution of $1$ degree.

\subsection*{Recovered Maps and Power Spectra}

\begin{figure*}[htb]
\includegraphics[scale=0.25]{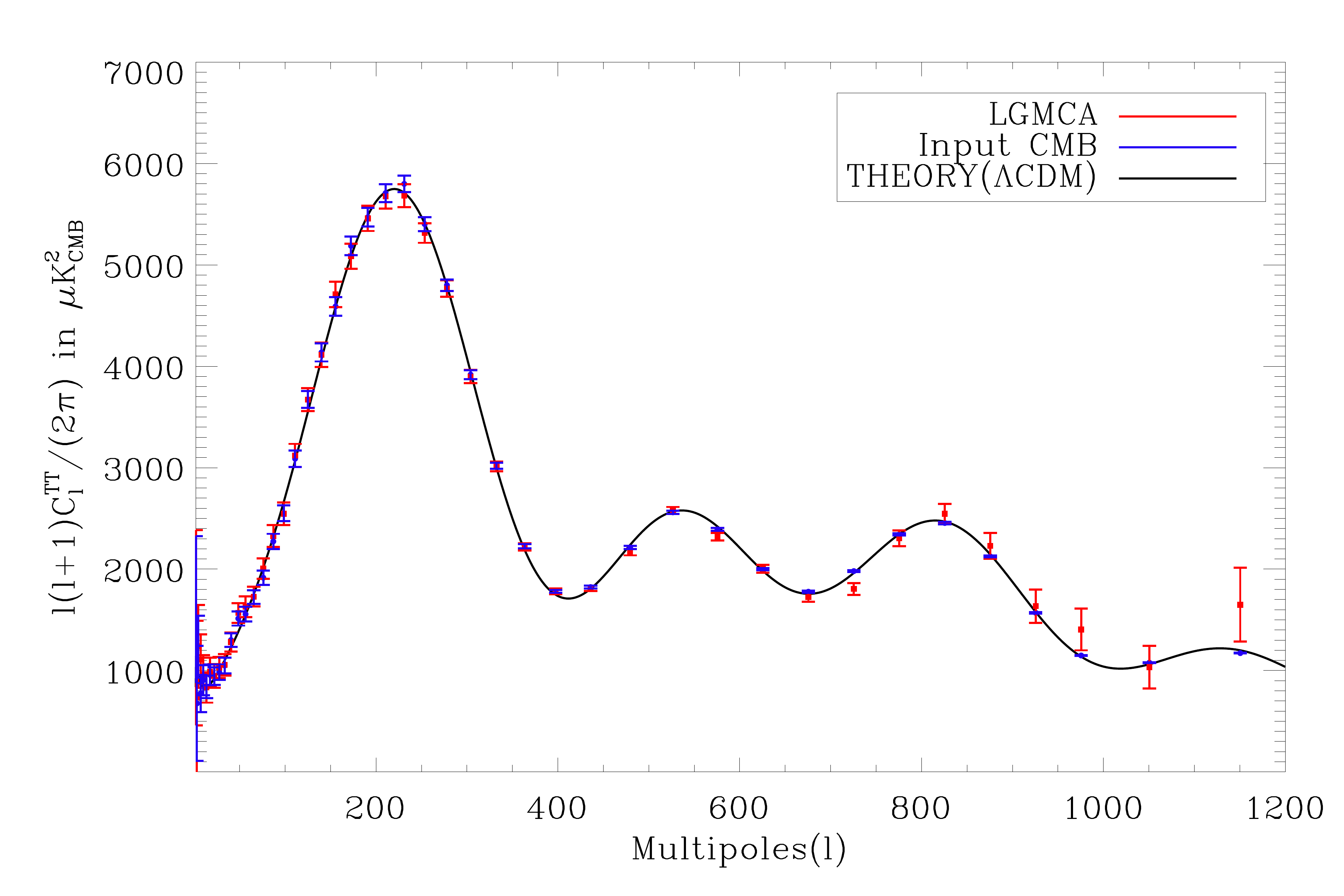}
\includegraphics[scale=0.25]{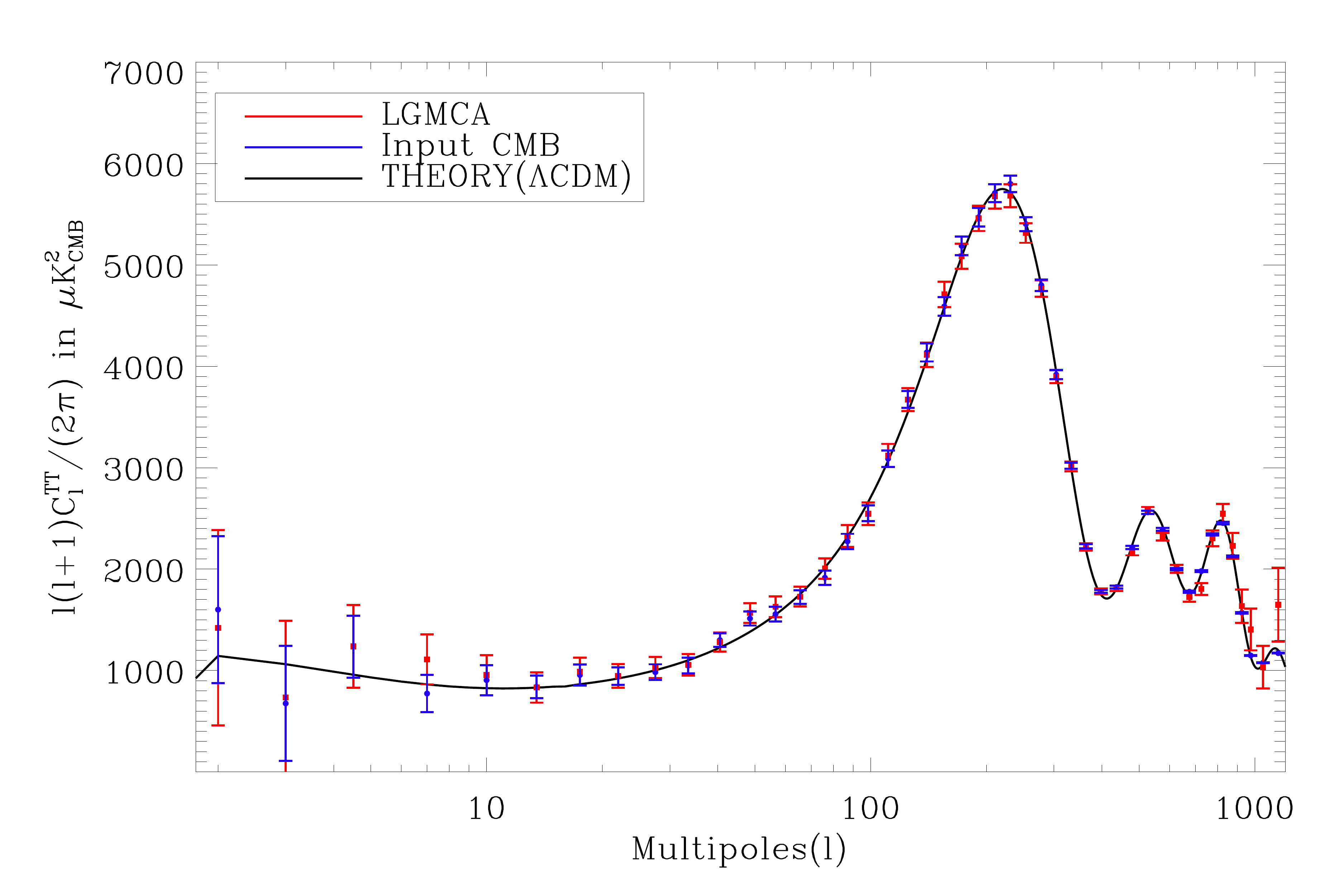}
\caption{Estimated CMB map power spectrum from simulated WMAP ($9$ years).}
\label{fig:simu_ps_1}
\end{figure*}

\begin{figure*}[tb]
\includegraphics[scale=0.25]{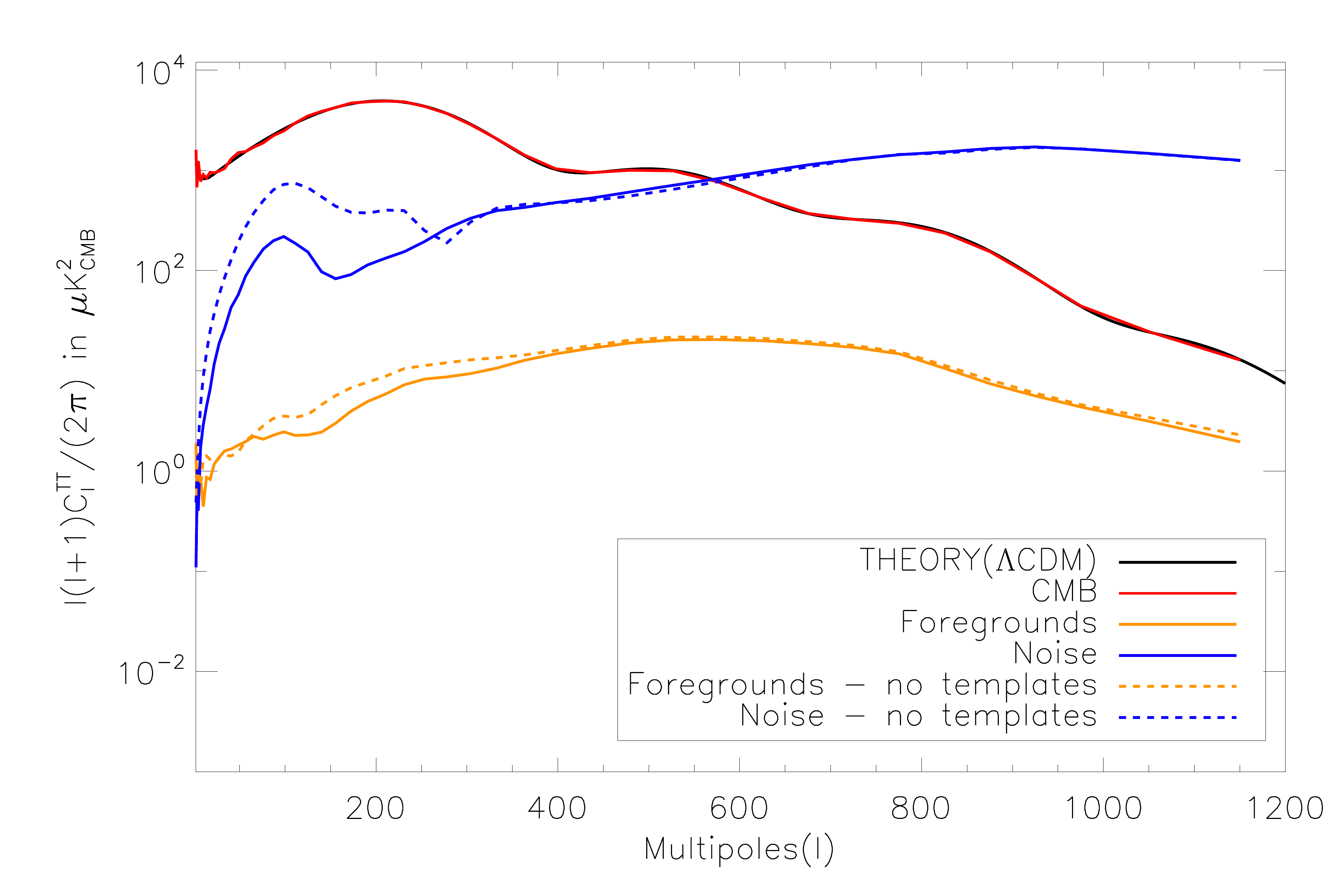}
\includegraphics[scale=0.25]{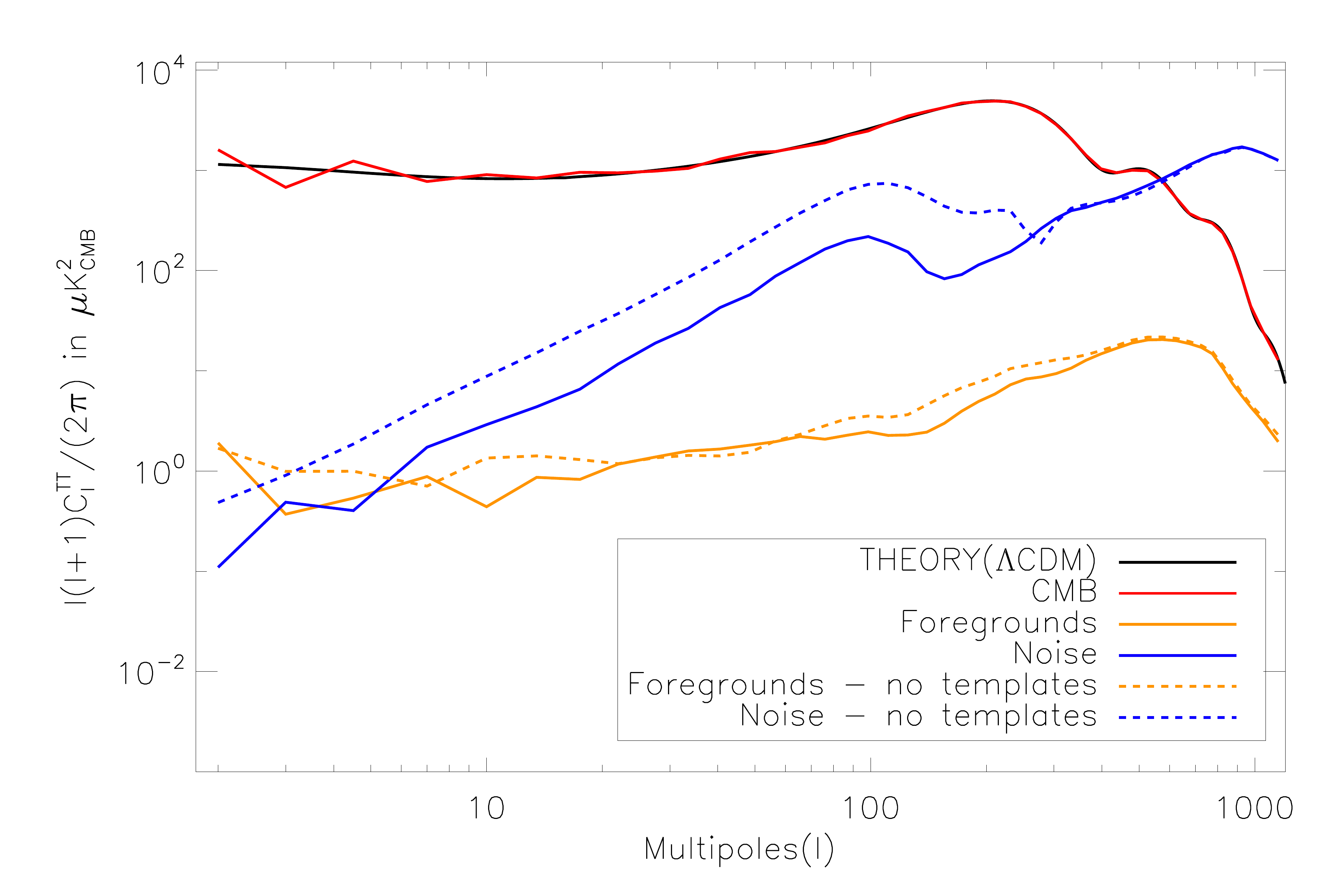}
\caption{Estimated CMB map, noise and remaining foregrounds power spectra from simulated WMAP ($9$ years) at $15$ arcmin resolution.}
\label{fig:simu_propa}
\end{figure*}

The power-spectrum is computed following the procedure described in Section~\ref{sec:mapest}. Figure~\ref{fig:simu_ps_1} displays the theoretical power-spectrum in black and the LGMCA estimated power-spectrum in red. The pseudo-spectrum of the input map is shown in blue; these points would correspond to a perfect estimation of the CMB map where only cosmic variance is a source of uncertainty. The larger $1\,\sigma$ red errors originate from the error coming from the remaining instrumental noise. In this experiment, $75\%$ of the sky coverage is used; the mask we used is a combination of point sources and galactic masks. These two plots show that the power-spectrum of the CMB map estimated after component separation does not show any statistically significant bias.\\
The use of simulations allow for a precise decomposition of the CMB estimation error into its different components~: CMB, remaining instrumental noise and foregrounds. For that purpose we apply the inversion parameters estimated with LGMCA independently to the simulated foregrounds and the instrumental noise. The resulting maps give the exact level of contamination of the CMB estimated by LGMCA. Figure~\ref{fig:simu_propa} shows the power spectra of the CMB as well as the residual noise and foregrounds that contaminate the estimated map. The resolution of the map estimated with LGMCA is $15$ arcmin; therefore the different spectra in Figure~\ref{fig:simu_propa} remains at the same resolution and are not deconvolved to infinite resolution. Again, exactly the same sky coverage of $75 \%$ is used in this experiment thus quantifying the exact level of foreground contamination of the estimated CMB power spectrum displayed in Figure~\ref{fig:simu_ps_1}. The two panels of Figure~\ref{fig:simu_propa} first show that the main source of contamination is the remaining instrumental noise which predominates for $\ell > 600$. This translates to the large error bars of the estimated power spectrum at small scales in Figure~\ref{fig:simu_ps_1}. For very low-$\ell$ ($\ell < 20$), the contribution of both the remaining noise and foregrounds is less than $1 \%$ which is way below the error related to the cosmic variance. In this experiment, the level of foreground contamination seem to be below $1 \%$ at all scales. This very low level has to be tempered~: the ancillary data, namely the composite all-sky H-alpha map of \citep{2003ApJS..146..407F} and the Finkbeiner thermal dust template \citep{Fink99} are also used within the PSM to produce the simulations of the free-free emission and thermal dust emission respectively. We should therefore expect the level of residual foregrounds to be higher when LGMCA is applied to the real WMAP data. Interestingly, we also applied LGMCA with exactly the same parameters except that only the WMAP maps without ancillary data were used. The contamination levels are featured in Figure~\ref{fig:simu_propa} in dashed line. If using templates indeed lowers the level of remaining foregrounds, their contribution is still much lower than the level of CMB. Noise is therefore the main source of contamination in the final CMB estimate whether external templates are used or not.

\begin{figure*}[htb]
\includegraphics[scale=0.25]{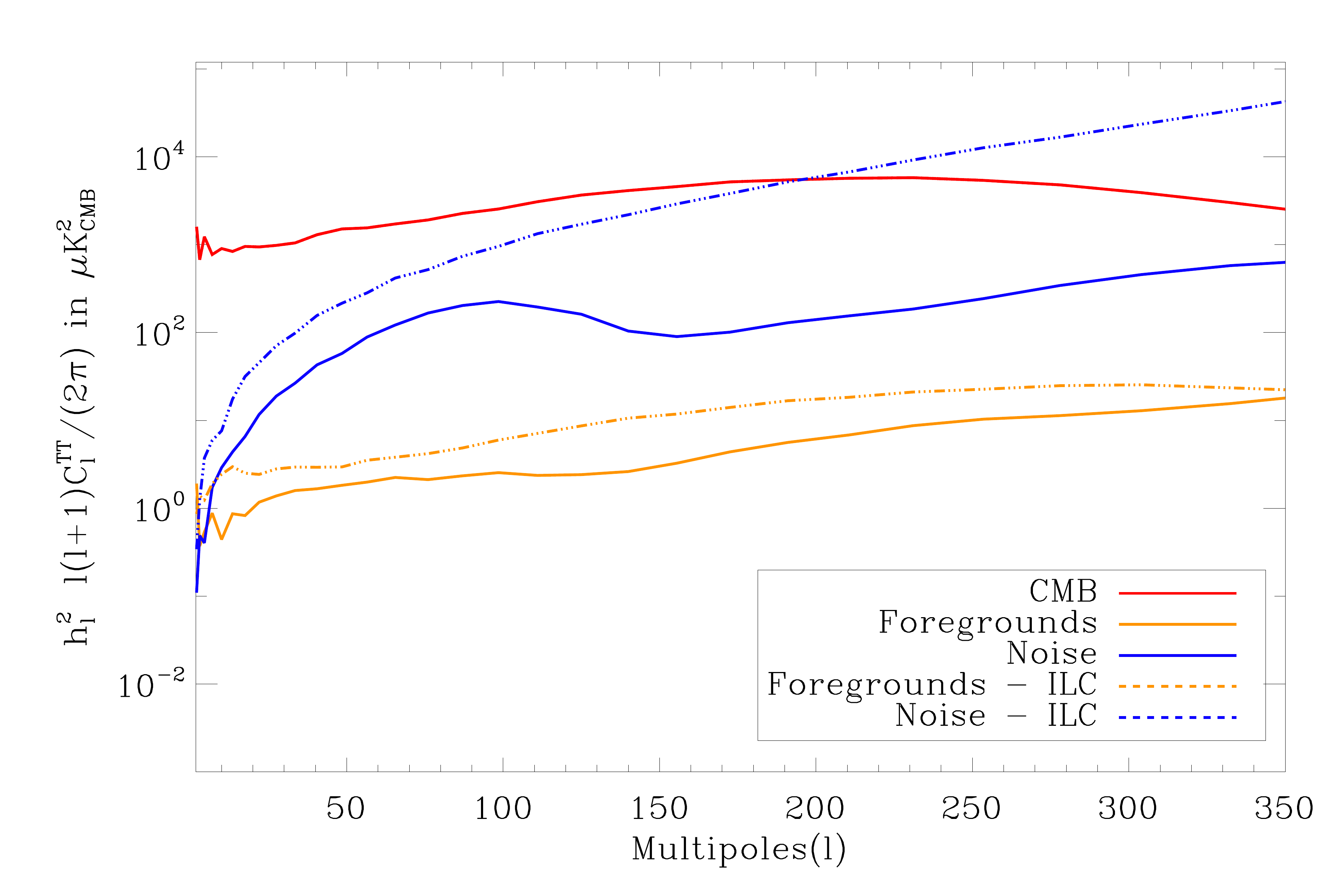}
\includegraphics[scale=0.25]{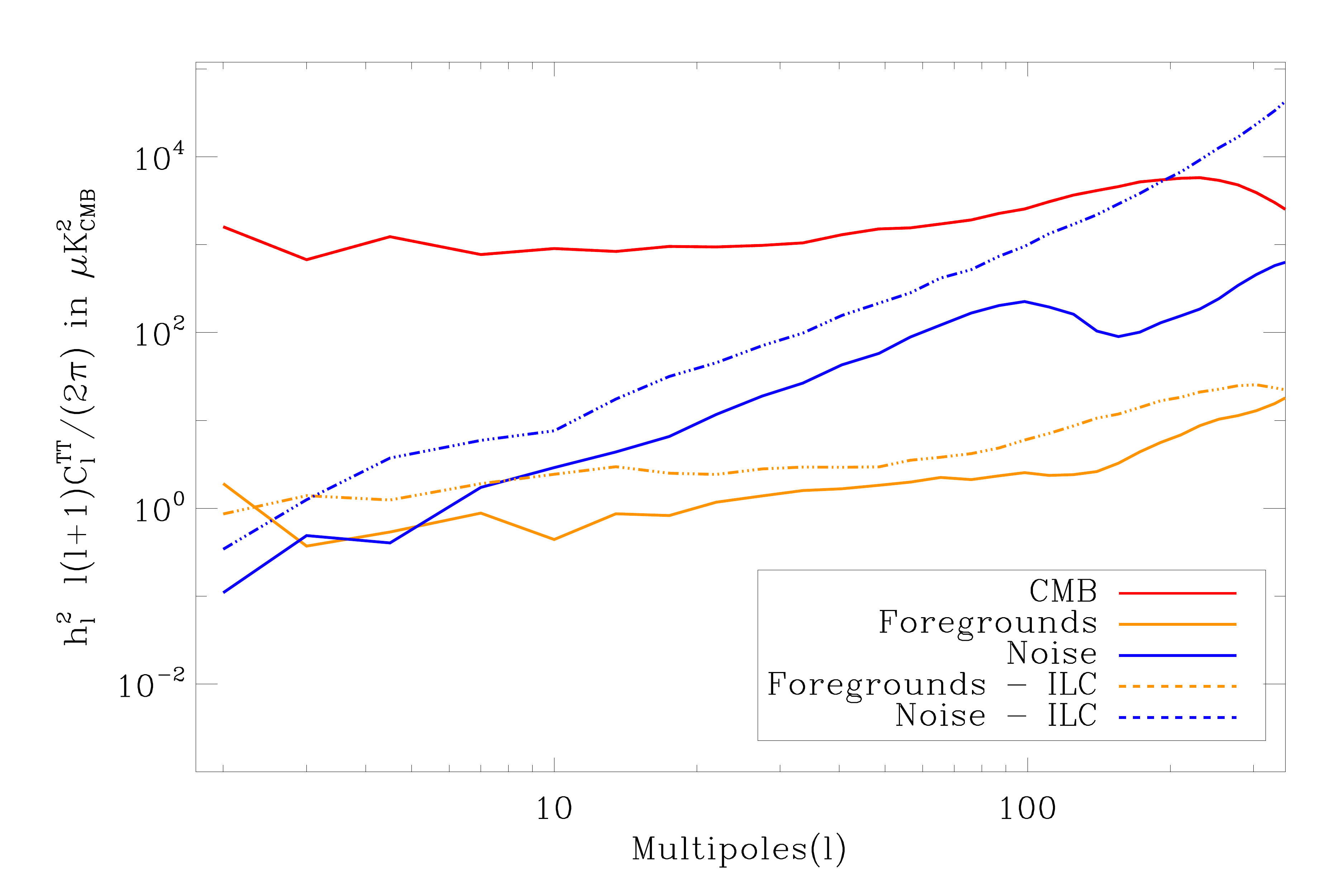}
\caption{Comparisons between LGMCA and ILC at $1$ degree resolution. Estimated CMB map, noise and remaining foregrounds power spectra from simulated WMAP ($9$ years). Please note that the amplitudes have been further amplified by $h_\ell^2$, the square of the $1$ degree resolution Gaussian beam.}
\label{fig:simu_propa_ILC}
\end{figure*}

Finally, in Figure \ref{fig:simu_propa_ILC} all components were propagated using weights computed by ILC and LGMCA. This figure illustrates that LGMCA is more efficient at lowering noise levels (due to the high amplification of noise in the ILC map when the $1$ degree deconvolution is performed) and foregrounds contamination (due to localization and the use of templates) compared to the computed ILC map.

\subsection*{Higher Order Statistics --- Non-Gaussianities}

\begin{figure*}[!ht]
\begin{center}
\includegraphics[scale=0.4]{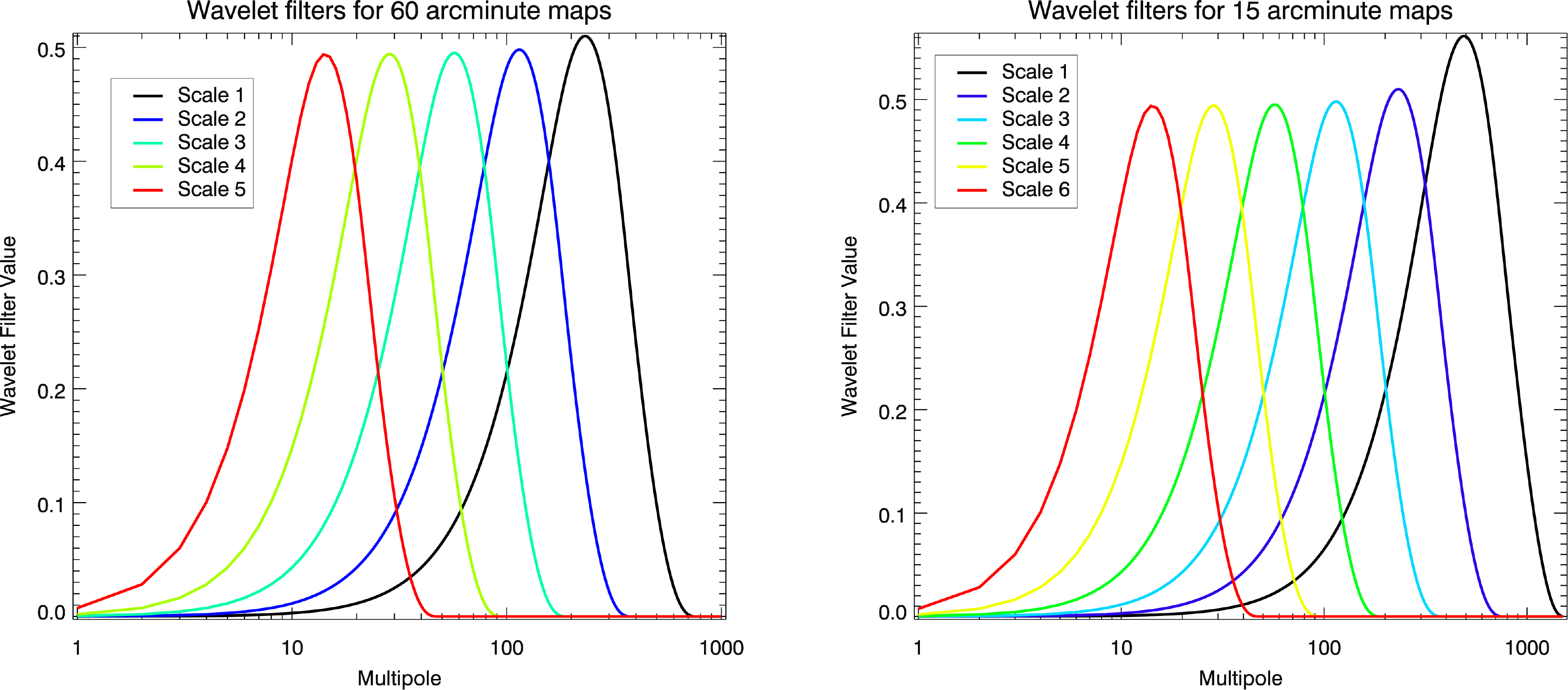}
\end{center}
\caption{Legendre coefficients of the wavelet filters employed for non-gaussianity analyses at $1$ degree (left) and $15$ arcminutes (right). These wavelets are well localized in pixel space, allowing a fine analysis per latitude bands.}
\label{fig:simu_hos_0}
\end{figure*}

The level of non-Gaussianity in the recovered CMB map provides a sanity check to measure and localize any remaining foreground contamination in the recovered CMB map, since in this case the CMB is generated as a Gaussian random field. In this work, we have computed the non-Gaussianity levels from the recovered CMB maps for LGMCA and ILC at $1$ degree, and for the LGMCA map at $15$ arcminutes. The $75\,\%$ mask was employed and sparse inpainting was performed to interpolate the signal inside the mask \citep{Starck2013}. 
The skewness and kurtosis were then computed on the simulation inside these masks on different wavelet scales using an isotropic undecimated wavelet on the sphere \citep{Starck05}, with the wavelet filters in spherical harmonic space described in Figure \ref{fig:simu_hos_0}. These statistics were then centered on the expected value (computed by propagating only the simulated noise and CMB) and normalized by the standard deviation computed from a set of $100$ CMB and noise realizations. These statistics were also computed at different latitude bands for each wavelet scale to assess the level of foreground contamination in the maps at various scales and positions. 

\begin{figure*}[hbt]
\begin{center}
\includegraphics[scale=0.3]{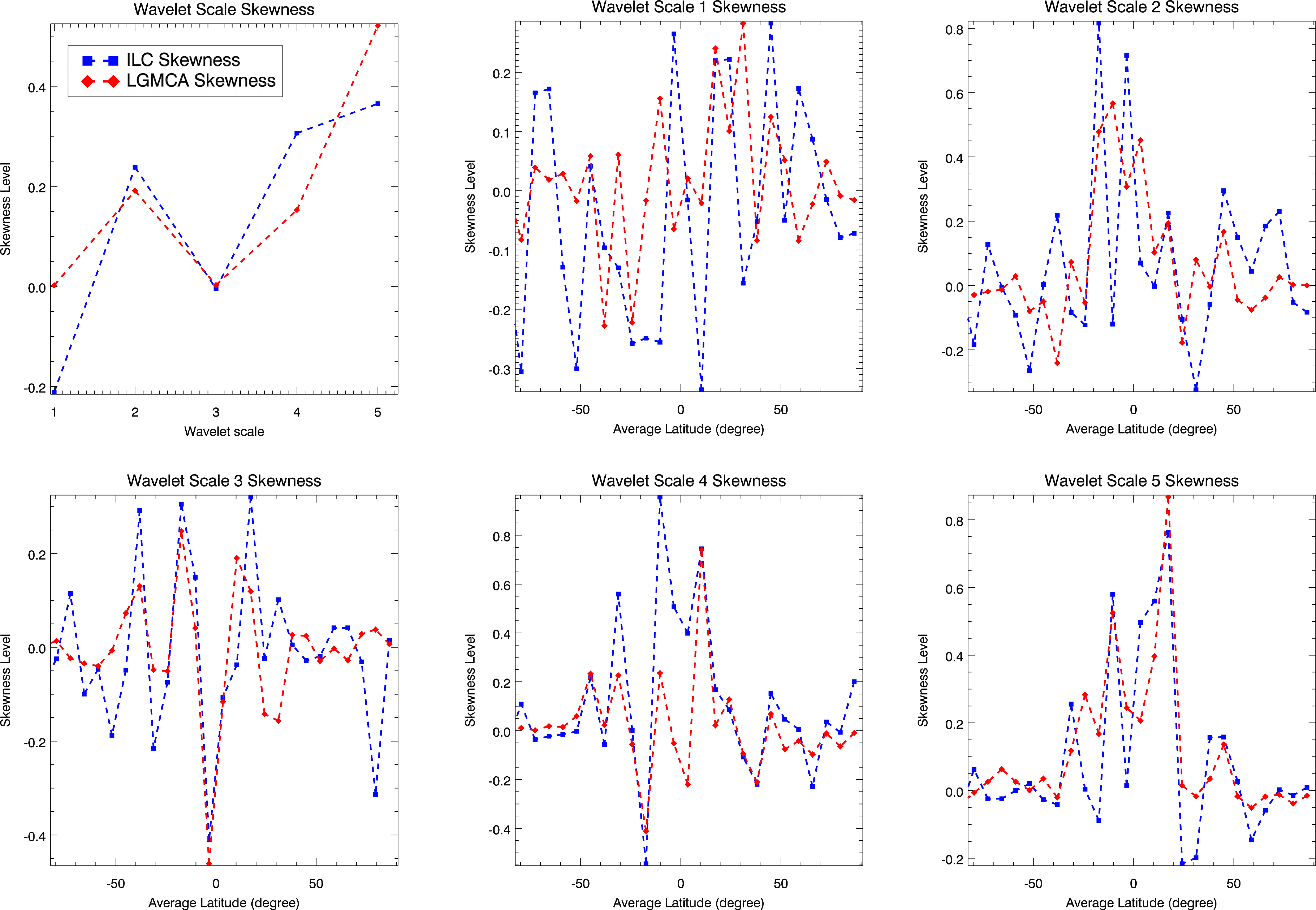}
\end{center}
\caption{Comparison of skewness levels in LGMCA (red) and ILC (blue) maps at 1 degree computed for various wavelet scales. These statistics were centered on the expected value and normalized from a set of $100$ simulations of CMB and Noise (see text). }
\label{fig:simu_hos_1}
\end{figure*}

\begin{figure*}[hbt]
\begin{center}
\includegraphics[scale=0.3]{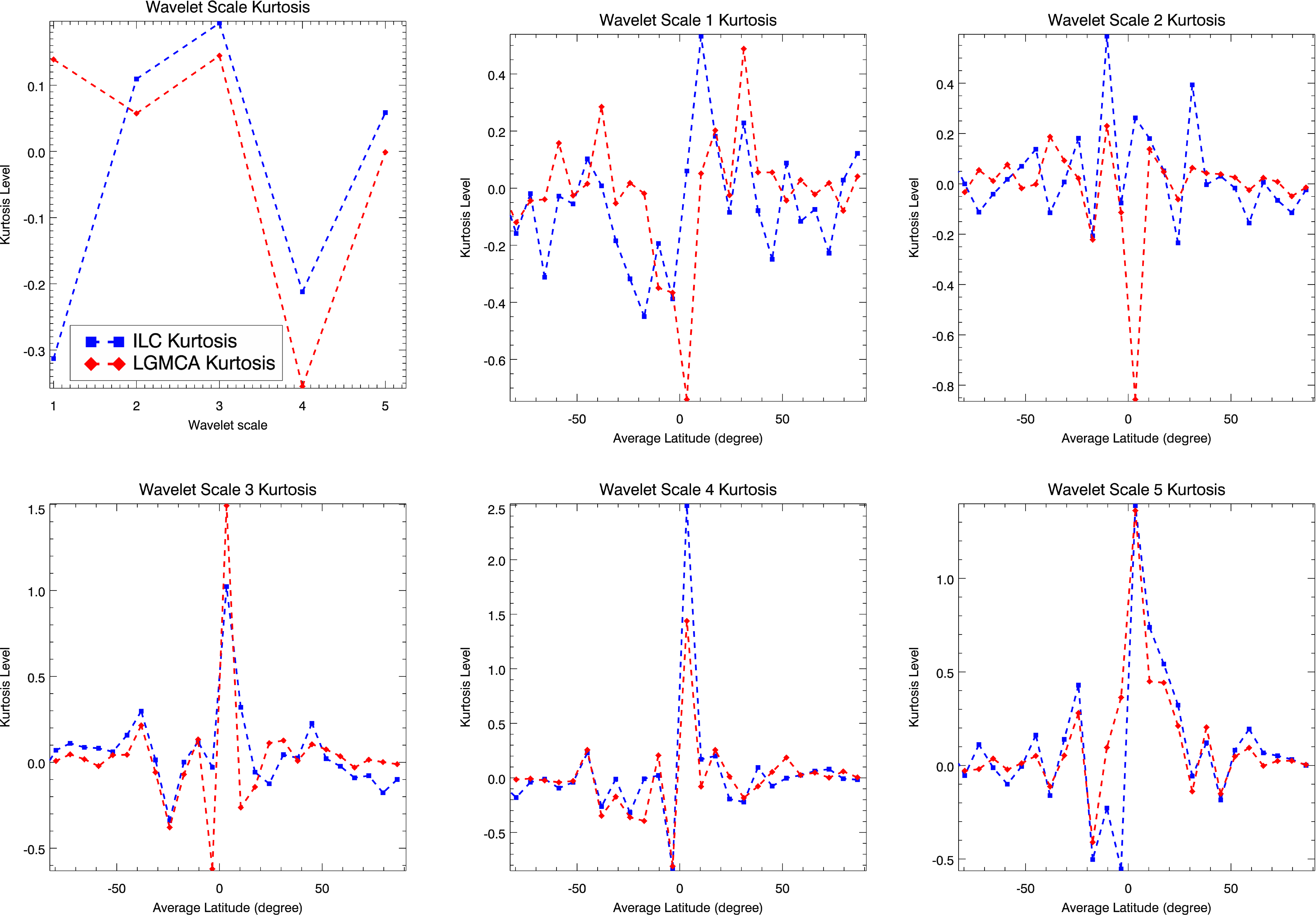}
\end{center}
\caption{Comparison of centered and normalized kurtosis in LGMCA (red) and ILC (blue) maps at $1$ degree computed for various wavelet scales. The same mask and set of simulations where employed as in Figure \ref{fig:simu_hos_1}.}
\label{fig:simu_hos_2}
\end{figure*}

\begin{figure*}[hbt]
\begin{center}
\includegraphics[scale=0.35]{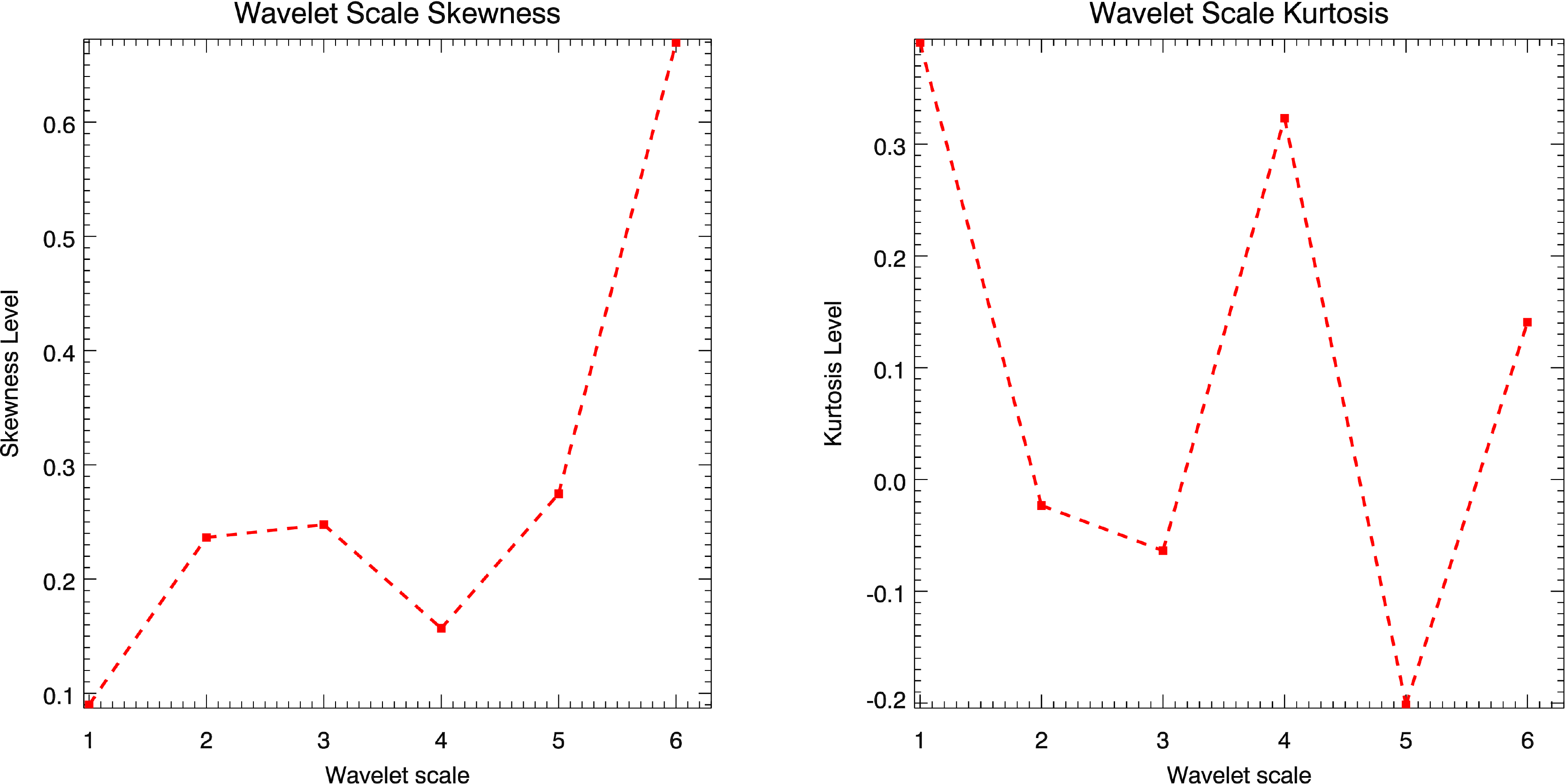}
\end{center}
\caption{Centered and normalized skewness and kurtosis  in LGMCA map at $15$ arcminutes computed for various wavelet scales. A $75 \%$ mask and a set of $100$ simulations of CMB and Noise were used to compute these statistics  (see text).  }
\label{fig:simu_hos_3}
\end{figure*}

\begin{figure*}[hbt]
\begin{center}
\includegraphics[scale=0.3]{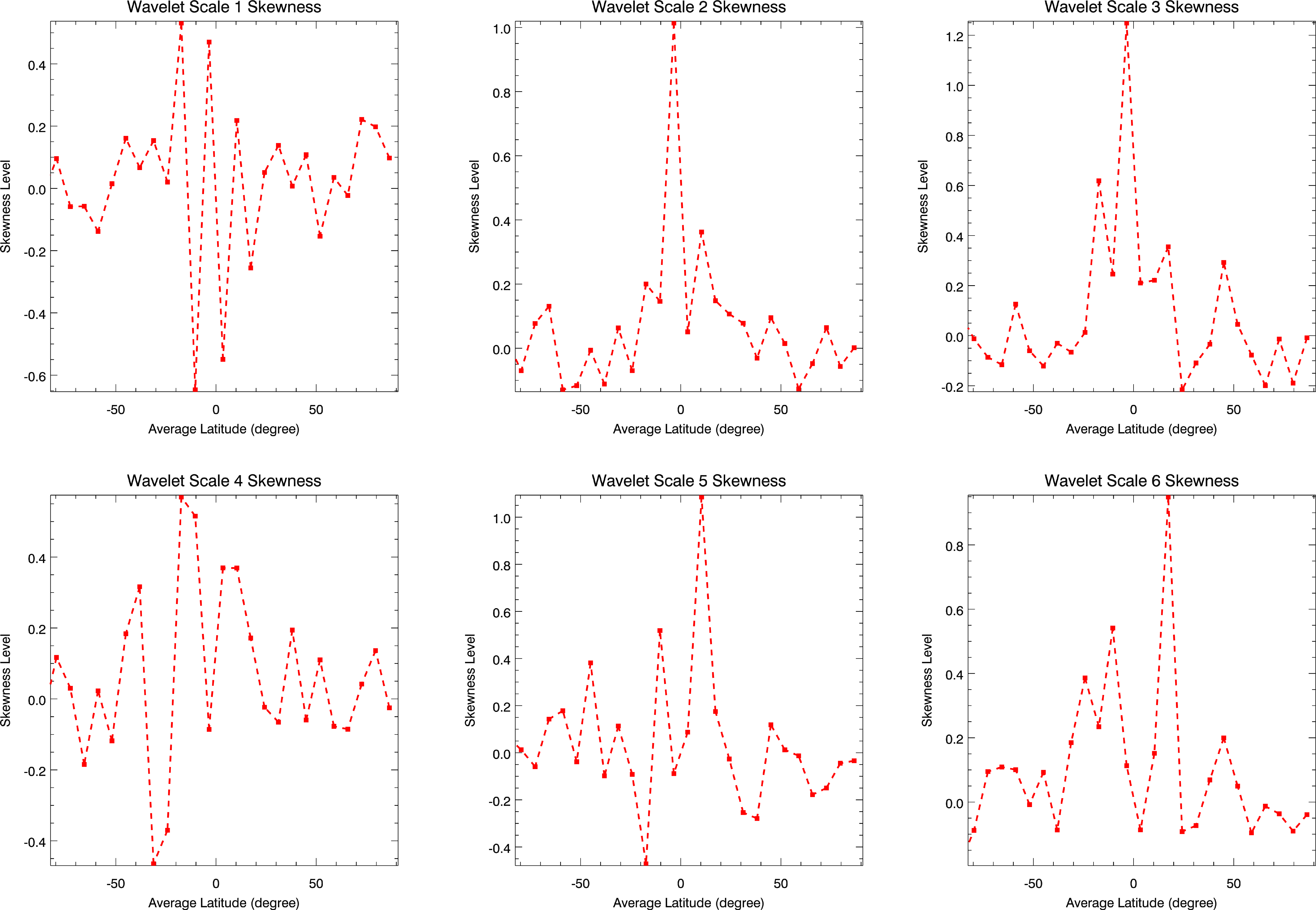}
\end{center}
\caption{Centered and normalized skewness in LGMCA map at $15$ arcminutes computed for various wavelet scales and location. The same mask and set of simulations were employed to derive the statistic as in Figure \ref{fig:simu_hos_3}. } 
\label{fig:simu_hos_4}
\end{figure*}

\begin{figure*}[hbt]
\begin{center}
\includegraphics[scale=0.3]{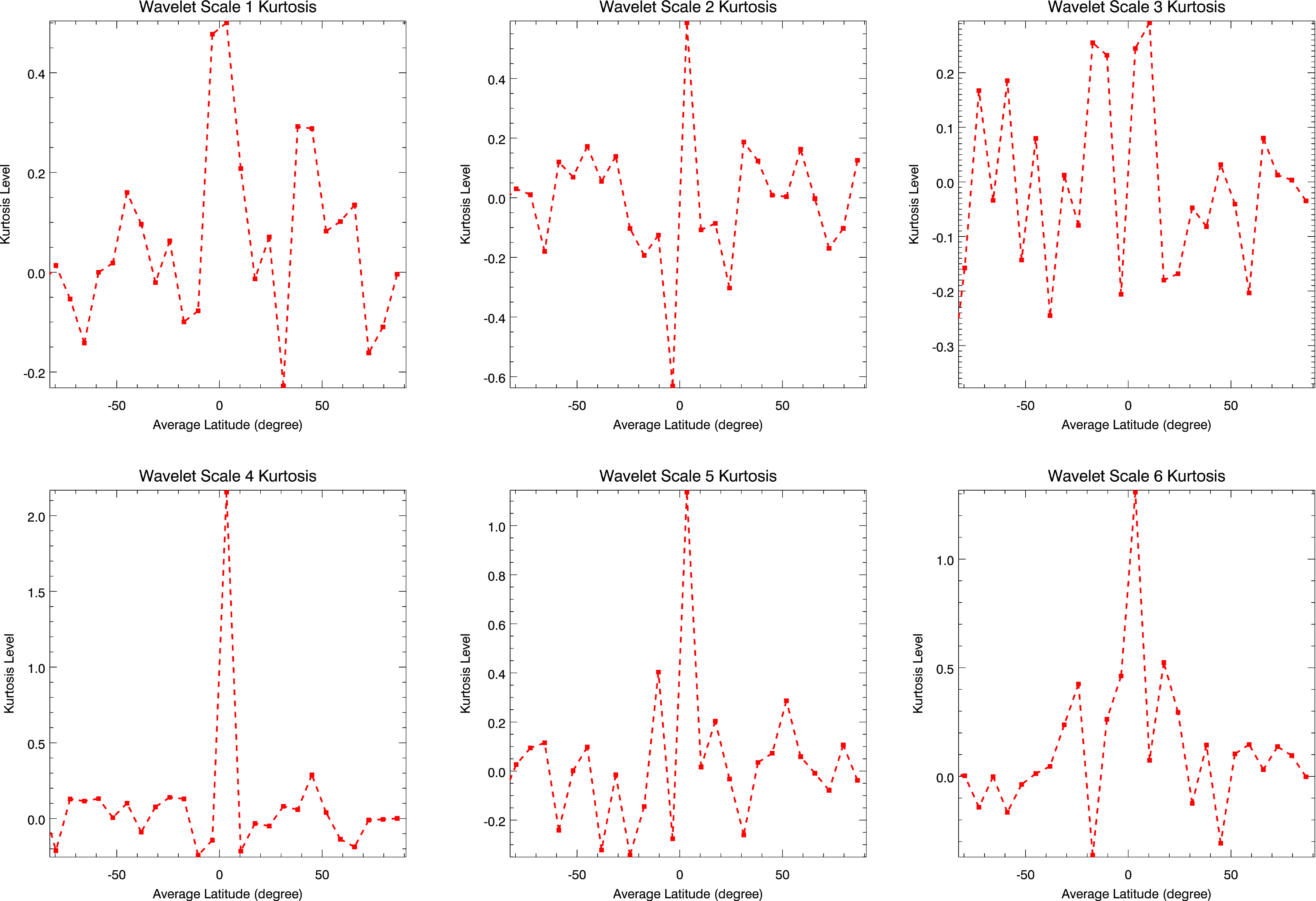}
\end{center}
\caption{Centered and normalized kurtosis in LGMCA map at $15$ arcminutes computed for various wavelet scales and location.The same mask and set of simulations were employed the statistic as in Figure \ref{fig:simu_hos_3}.}
\label{fig:simu_hos_5}
\end{figure*}

The comparison between LGMCA and ILC at $1$ degree is displayed  in Figures \ref{fig:simu_hos_1} and \ref{fig:simu_hos_2}. For both methods, the skewness and kurtosis are  compatible with the error bars due to propagated noise and cosmic variance, with a maximal detection at $2.5 \sigma$ close to the galactic center. The same tests were also performed for the LGMCA map at the full resolution of $15$ arcminutes and are displayed in Figures \ref{fig:simu_hos_3}, \ref{fig:simu_hos_4} and \ref{fig:simu_hos_5}.  The difference observed between the LGMCA non-gaussianity levels and those computed from the simulation without foregrounds is compatible with the errors expected at that resolution. 

\end{document}